\documentclass[11pt]{article}
\usepackage{jcappub}
\usepackage{dsfont}
\usepackage{bbm}
\usepackage{slashed}
\numberwithin{equation}{section}

\newcommand{\be}{\begin{equation}}
\newcommand{\ee}{\end{equation}}
\newcommand{\bea}{\begin{eqnarray}}
\newcommand{\eea}{\end{eqnarray}}

\newcommand{\vx}{\vec{x}}

\title{Tensor to scalar ratio and large scale power suppression from pre-slow roll initial conditions.}

\author{Louis Lello, }
\author{Daniel Boyanovsky}

\affiliation{Department of Physics and
Astronomy, University of Pittsburgh, Pittsburgh, PA 15260}

\emailAdd{lal81@pitt.edu}

\emailAdd{boyan@pitt.edu}

\abstract{We study the corrections to the power spectra of curvature and tensor perturbations and the tensor-to-scalar ratio $r$ in single field slow roll inflation with standard kinetic term due to initial conditions imprinted by a ``fast-roll'' stage prior to slow roll.   For a wide range of initial inflaton kinetic energy,  this stage lasts only a few e-folds and  merges smoothly with   slow-roll thereby leading to   non-Bunch-Davies initial conditions for modes that exit the Hubble radius during slow roll. We describe a program that yields the dynamics in the fast-roll stage while matching to the slow roll stage in a manner that is independent of the inflationary potentials. Corrections to the power spectra are encoded in a ``transfer function'' for initial conditions $\mathcal{T}_\alpha(k)$, $\mathcal{P}_{\alpha}(k)=P^{BD}_{\alpha}(k)\mathcal{T}_\alpha(k)$, implying a modification of the ``consistency condition'' for the tensor to scalar ratio at a pivot scale $k_0$: $r(k_0)= -8n_T(k_0)\,\big[ {\mathcal{T}_{T}(k_0)}/{\mathcal{T}_{\mathcal{R}}(k_0)} \big]$. We obtain  $\mathcal{T}_\alpha(k)$ to leading order in a Born approximation valid for modes of observational relevance today. A fit yields $\mathcal{T}_\alpha(k) =1+ A_{\alpha}k^{-p}\cos[2\pi \omega  k/H_{sr}+\varphi_\alpha]$, with $1.5 \lesssim p \lesssim 2$, $\omega \simeq 1$ and $H_{sr}$ the Hubble scale during slow roll inflation, where curvature and tensor perturbations feature the \emph{same} $p,\omega$ for a wide range of initial conditions. These corrections lead to both a suppression of the quadrupole and oscillatory features in both $P_R(k)$ and  $r(k_0)$ with a period of the order of the Hubble scale during slow roll inflation.  The results are quite general and independent of the specific inflationary potentials, depending solely on the ratio of kinetic to potential energy $\kappa$ and the slow roll parameters $\epsilon_V,\eta_V$ to leading order in slow roll. For a wide range of $\kappa$ and the values of $\epsilon_V;\eta_V$ corresponding to the upper bounds from Planck, we find that the low quadrupole is consistent with the results from Planck, and the   oscillations   in $r(k_0)$ as a function of $k_0$ \emph{could} be observable  if the modes corresponding to the quadrupole and the pivot scale crossed the Hubble radius very few  ($2-3$) e-folds after the onset of slow roll. We comment on possible impact on the recent BICEP2 results.}

\arxivnumber{}
\begin{document}

\maketitle

\section{Introduction}
Inflation not only  provides a solution to the horizon and flatness problems but also furnishes  a mechanism for generating scalar (curvature) and tensor (gravitational wave)   quantum fluctuations\cite{staro2,guth,linde,al}. These fluctuations seed the small temperature inhomogeneities in the cosmic microwave background (CMB) upon reentering the particle horizon during recombination. Most inflationary scenarios    predict  a  nearly gaussian and nearly scale invariant power spectrum of adiabatic fluctuations \cite{mukh,kolb,riotto2,baumann,giov}. These important predictions of inflationary cosmology are supported by observations of the cosmic microwave background\cite{wmap7,wmap9,planck,planck2} which are beginning to discriminate among different scenarios.

Recent results from the Planck collaboration\cite{planck,planck2,planck3} have provided the most precise analysis of the (CMB) to date, confirming the main features of the inflationary paradigm, but at the same time highlighting perplexing large scale anomalies, some of them, such as a low quadrupole,  dating back to the early observations of the Cosmic Background Explorer (COBE)\cite{cobe,bondlow}, confirmed with greater accuracy by WMAP\cite{wmaplow} and Planck\cite{planck,planck2,planck3}. Recently the BICEP2 collaboration \cite{bicep2} has provided the first measurement of primordial B-waves, possibly the first direct evidence of inflation.

The interpretation and statistical significance of these anomalies is a matter of much debate, but being associated with the largest scales, hence the most primordial aspects of the power spectrum, their observational evidence is not completely dismissed\cite{copi}. The possible origin of the large scale anomalies is vigorously discussed, whether these are of primordial origin or a consequence of the statistical analysis (masking) or secondary anisotropies is still an open question. Some studies claim the removal of    large scale anomalies (including the suppression of power of the low multipoles) after substraction of the integrated Sachs-Wolfe effect (ISW)\cite{francis}, however a different recent  analysis\cite{rassat} finds that the low quadrupole becomes even \emph{more} anomalous after subtraction of the (ISW) contribution, although some expansion histories may lead to
an (ISW) suppression of the power spectrum\cite{tarun}. The most recent Planck\cite{planck,planck2,planck3} data still finds  a statistically significant discrepancy at low multipoles, reporting a power deficit $5-10\%$ at $l \lesssim 40$ with $2.5-3\,\sigma$ significance. This puzzling and persistent result stands out in an otherwise consistent picture of $\Lambda CDM$ insofar as the (CMB) power spectrum is concerned. Recent analysis of this lack of power at low $l$\cite{grup} and large angles\cite{copi}, suggests that while limited by cosmic variance,  the possibility  of the primordial origin of the large scale anomalies cannot be dismissed and  merits further study.

The simpler inflationary paradigm that successfully explains the cosmological data relies on the dynamics of a scalar field, the inflaton, evolving slowly during the inflationary stage with the dynamics determined by a fairly flat potential. This simple, yet observationally supported inflationary scenario is referred to as slow-roll inflation\cite{kolb,mukh,riotto2,baumann,giov}. Within this   scenario   wave vectors of cosmological relevance    cross the Hubble radius during inflation with nearly constant amplitude leading to a nearly scale invariant power spectrum. The quantization of the gaussian fluctuations (curvature and tensor) is carried out by imposing a set of initial conditions so that fluctuations with wavevectors deep inside the Hubble radius are described by Minkowski space-time free field mode functions. These are known as Bunch-Davies initial conditions\cite{bunch} (see for example\cite{kolb,mukh,baumann,giov} and references therein).

The issue of modifications of these initial conditions and the potential impact on the inflationary power spectra\cite{ini1,ini2,ini3,ini4,ini5,holini1,holini2,martin,daniels,picon,staro4},  enhancements to non-gaussianity\cite{holtol1,holtol2,ganc,parker1,parker2,porto,dustin,amjad,kundu,jain,meerburg}, and   large scale structure\cite{ganckoma} have been discussed in the literature. Furthermore, arguments presented in  refs.\cite{picon,mottola}   suggest that Bunch-Davies initial conditions are not the most natural ones to consider and may be unstable to small perturbations.

Whereas the recent results from Planck\cite{planck,planck2,planck3} provide tight constraints on primordial non-gaussianities including modifications from initial conditions, these constraints \emph{per se} do not apply directly to the issue of initial conditions on other observational aspects.

Non-Bunch-Davies initial conditions arising from a pre-slow roll stage during which the (single) inflaton field features   ``\emph{fast-roll}'' dynamics have been proposed as a \emph{possible} explanation of power suppression at large scales\cite{lindefast,contaldi,boyan3,hectordestri,nicholson,reviunos,schwarz1,schwarz2}. This kinetically dominated fast roll stage has been considered previously using the Hamilton-Jacobi form of the Friedman equations in ref.\cite{kinney1} with some of the consequences due to this type of scalar driven cosmology examined in detail in\cite{kinney2, woodard1, woodard2}. Alternative pre-slow-roll descriptions in terms of interpolating scale factors pre (and post) inflation have also been discussed in ref.\cite{parkglenz}. The influence of non-Bunch Davies initial conditions arising from a fast-roll stage just prior to slow roll on the infrared aspects of nearly massless scalar fields in de Sitter space time have been studied in ref.\cite{lellodS}.  Recent work \cite{lasenby} has shown that a kinetically dominated regime is in fact quite a generic feature under a very broad class of single field inflationary models providing further incentive for consideration of fast roll scenarios.

\vspace{2mm}

\textbf{Motivations, goals and summary of results:}

Inflationary scenarios predict the generation of primordial gravitational waves and their detection remains one of the very important goals of observational cosmology. Planck\cite{planck} has placed constraints on the tensor to scalar ratio of $r < 0.11\,(95\% CL)$ while the BICEP \cite{bicep2} experiment has recently reported a measurement of $r=0.20^{+0.07}_{-0.05}$. The BICEP value is much larger than many had expected and there exist models which can generate enhancements, refs \cite{mazumdar2} for example, which could explain the largeness of this value.

Suggestions of how to relieve the tension between the two experiments have been put forth where a possible solution invokes a running of the spectral index. Recently, in ref.\cite{contaldi2}, a comparison between models featuring a running spectral index and models with a large scale power suppression has been made with the aim of determining which model relieved the tension most effectively. It was shown in this reference that a large scale power suppression of ~35\% yielded a considerably better fit to data than allowing a running of the spectral index, further improving the claims that the low l anomaly should be taken seriously.

The high amount of tension between these two experiments may be alleviated by future and forthcoming observations that will continue to constrain this important quantity, a quantity from which ultimately the scale of inflation may be extracted\cite{planck},
\be V_{inf} =(1.94\times 10^{16}\,\mathrm{GeV})^4 \,\Big(\frac{r}{0.12}\Big)\,. \label{inflascale} \ee
 A distinct prediction of single field slow-roll inflationary models with a standard kinetic term is
\be r= 16\epsilon_V = -8 n_T \label{predi}\ee with $\epsilon_V$ a (potential) slow roll parameter and $n_T$ is the index of the power spectrum of gravitational waves. The relation (\ref{predi}) is often quoted as a ``consistency relation''. This relation is obtained by imposing Bunch-Davies initial conditions on tensor perturbations during the near de Sitter slow roll stage\cite{kolb,mukh,baumann,giov}.

Our goals in this article are the following:
 \begin{itemize}
  \item{Motivated by the recent results from the PLANCK collaboration\cite{planck,planck2,planck3} reporting the persistence of anomalies for small $l$ and large angular scales we study
  the modifications to the power spectra of curvature and tensor perturbations and the tensor  to scalar ratio  arising from non-Bunch Davies initial conditions imprinted from a pre-slow roll stage in which the dynamics of the scalar field is dominated by the kinetic term, namely a ``fast-roll stage''.
       While previous studies of modifications of the scalar power spectrum from a fast roll stage focused on specific realizations of the inflationary potential, our goal is to extract the main corrections   \emph{without resorting to a specific choice of the potential} but by parametrizing the fast roll stage by the initial ratio of kinetic to potential energy of the inflaton, $\dot{\Phi}^2_i/2V = \kappa$, and the potential slow roll parameters $\epsilon_V,\eta_V$ which have been constrained by Planck and WMAP-polarization (Planck+WP)\cite{planck} to be $\epsilon_V < 0.008 ~ (95\% \,CL);\eta_V = -0.010^{+0.005}_{-0.011}$.}

        \item{To explore  possible \emph{correlations} between the suppression of the low multipoles in the temperature power spectrum, and features in the tensor to scalar ratio $r(k_0)$ as a function of the pivot scale, and more generally,  to the  power spectrum of tensor perturbations,   as a consequence of the fast roll stage.}

    \item{To assess the scales and general aspects of features in the power spectra resulting from the modification of the initial conditions and their potential observability. }

            \end{itemize}

            \vspace{2mm}

\textbf{Brief summary of results:}
A fast roll stage prior to slow roll leads to non-Bunch-Davies conditions on the observationally relevant mode functions that cross the Hubble radius during slow roll. These modifications yield oscillatory corrections to the power spectra of curvature and tensor perturbations, with a period determined by the Hubble scale during slow roll inflation, and a modification of the consistency condition for the tensor to scalar ratio $r$ with oscillatory features as a function of the pivot scale. The results are general and do not depend on the specific form of the inflationary potential but to leading order in slow roll depend only on $\kappa;\epsilon_V;\eta_V$. We describe a systematic program that yields the solution interpolating between the fast and slow roll stages based on a derivative expansion and separation of scales, which is
\emph{independent of the inflationary potentials} provided these are monotonic and can be described in a derivative expansion characterized by slow roll parameters. The Non-Bunch Davies initial conditions from the fast roll stage lead to corrections to the power spectra of scalar and tensor perturbations in the form of oscillatory features with a typical frequency determined by the Hubble scale during slow roll.  The corrections to the power spectrum for curvature perturbations lead to a suppression of the quadrupole that is correlated with the  oscillatory features in the tensor to scalar ratio $r(k_0)$ as a function of the pivot scale $k_0$. The quadrupole suppression is consistent with the latest results from Planck\cite{planck2} and the oscillatory features in $r(k_0)$  \emph{could} be observable\cite{meersper} \emph{if} the mode corresponding to the Hubble radius today crossed the Hubble radius a few e-folds from the beginning of slow roll.

  \section{Fast roll stage:}
 We consider a spatially flat Friedmann-Robertson-Walker (FRW) cosmology with
\be ds^2 = dt^2-a^2(t)(d\vec x)^2 =
C^2(\eta)[d\eta^2 - (d\vec x)^2]~~;~~C(\eta) \equiv a(t(\eta)) \; , \ee where $t$ and $\eta$ stand
for cosmic and conformal time respectively and consider   curvature and tensor perturbations. The dynamics of the scale factor in single field inflation is determined by Friedmann and covariant conservation equations
\be H^2 = \frac{1}{3M^2_{Pl}}\Bigg[\frac{1}{2} \dot{\Phi}^2 + V(\Phi)\Bigg]~~;~~ \ddot{\Phi}+3H\dot{\Phi}+V'(\Phi) = 0\,. \label{inflaton}\ee
During the slow roll near de Sitter stage,
\be H^2_{sr} \simeq  \frac{V_{sr}(\Phi)}{3M^2_{Pl}} ~~;~~  3H\dot{\Phi}+V'_{sr}(\Phi) \simeq  0\,. \label{slowly}\ee This stage is   characterized by the smallness of the (potential) slow roll parameters\cite{mukh,kolb,riotto2,baumann,giov}
\be \epsilon_V = \frac{M^2_{Pl}}2 \;
\left[\frac{V'_{sr}(\Phi )}{V_{sr}(\Phi )} \right]^2  \simeq \frac{\dot{\Phi}^2_{sr}}{2M^2_{Pl}H^2}    \quad , \quad
\eta_V = M^2_{Pl} \; \frac{V''_{sr}(\Phi)}{V_{sr}(\Phi)}   \; , \label{slowroll}
\ee (here $M_{Pl}=1/\sqrt{8\pi\,G}$ is the \emph{reduced} Planck mass).

Instead, in this section we consider   an initial stage dominated by the kinetic term, namely a fast roll stage,  thereby neglecting the term $V'$ in the equation of motion for the inflaton,  (\ref{inflaton}) and consider the potential to be (nearly) constant and equal to the potential during the slow roll stage, namely $V(\Phi) \simeq V(\Phi_{sr})\equiv V_{sr}$. In the following section we relax this condition in a consistent expansion in $\sqrt{\epsilon_V}$.
\bea && H^2    =  \Big(\frac{\dot{a}}{a}\Big)^2 = \frac{1}{3M^2_{Pl}}\Bigg[\frac{1}{2} \dot{\Phi}^2 + V_{sr} \Bigg]\label{FRW2}\\ && \ddot{\Phi}+3H\dot{\Phi}  \simeq 0\,. \label{inflaton2}\eea   The solution to (\ref{inflaton2}) is given by
\be \dot{\Phi}(t)= \dot{\Phi}_i \Big(\frac{a_i}{a(t)}\Big)^3 \,, \label{solinfla}\ee an initial value of the velocity damps out and the slow roll stage begins when $\ddot{\Phi} \ll 3H_{sr}\dot{\Phi}\simeq -V'_{sr}(\Phi)$. During the slow roll stage when $3H_{sr}\dot{\Phi}_{sr} \simeq -V'_{sr}$ it follows that
\be   \frac{3\dot{\Phi}^2_{sr}}{2V_{sr} }  = \epsilon_V \,.\label{fidotsr}\ee

The dynamics enters the slow roll stage when $\dot{\Phi} \sim \mathcal{O}(\sqrt{\epsilon_V})$ as seen by (\ref{slowroll}). To a first approximation, we will assume that Eq.(\ref{solinfla}) holds not only for the kinetically dominated epoch, but also until the beginning of slow roll ($\dot{\Phi}^2 \sim \epsilon_V$). In Section \ref{sec:match}, this approximation is justified and the error incurred from such an assumption is made explicit. The dynamics enters the slow roll stage at a value of the scale factor $a(t_{sr})\equiv a_{sr}$ so that
\be \dot{\Phi}_{sr} a^3_{sr} = \dot{\Phi}_i a^3_i \,.\label{equ}\ee  We now use the freedom to rescale the scale factor to set
 \be a(t_{sr}) = a_{sr} = 1\,, \label{choice}\ee this normalization is particularly convenient to establish
 when a particular mode crosses the Hubble radius during slow roll, an important assessment in the analysis below.

 In terms of these definitions and eqn. (\ref{equ}), we have that during the fast roll stage
\be \dot{\Phi}(t) =\frac{\dot{\Phi}_{sr} }{ {a}^3(t)}   \,. \label{fit}\ee Introducing
\be H^2_{sr} \equiv \frac{V_{sr} }{3M^2_{Pl}}\,, \label{Hsr}\ee Friedmann's equation becomes
\be \frac{\dot{ {a}}(t)}{a(t)} = H_{sr} \Bigg[1+\frac{\epsilon_V}{3\, {a}^6(t)}\Bigg]^{1/2} \,. \label{frie}\ee This equation for the scale factor can be readily integrated to yield the solution
\be  {a}(t) = \Bigg[\Bigg( \frac{\epsilon_V}{3}\Bigg)^{1/2} \sinh[\theta(t)] \Bigg]^{1/3}~~;~~\theta(t)=\theta_0+3H_{sr}t \label{tila}\ee where $\theta_0$ is an integration constant chosen to be
\be e^{-\theta_0} = \sqrt{\frac{\epsilon_V}{12}} \,,\label{teta0} \ee so that at long time $a(t) = e^{H_{sr}t}$. The slow roll stage begins when $a(t_{sr}) =1$ which corresponds
   to the value of $\theta_{sr}=\theta(t_{sr})$ given by
\be e^{-\theta_{sr}}= f\big(\frac{\epsilon_V}{3}\big) \label{thetasr} \ee where to simplify notation later we defined
\be f(s) = \frac{\sqrt{s}}{1+\sqrt{1+s}}\,. \label{fofs}\ee Introducing the dimensionless ratio of kinetic to potential contributions at the initial time $t_i$
 \be \frac{\dot{\Phi}^2_i}{2V_{sr}} = \kappa \label{kapa} \,, \ee and \emph{assuming} that the potential does not vary very much between the initial time and the onset of  slow roll (this is quantified below),   it follows from (\ref{equ})  that
\be   a^6_i = \frac{\dot{\Phi}^2_{sr}}{2V_{sr}\kappa} = \frac{\epsilon_V}{3\kappa} \label{equa2}\ee where we have used (\ref{fidotsr}). Combining this result with (\ref{tila}), we find that at the initial time $\theta_i=\theta(t_i)$ is given by
\be e^{-\theta_i} = f(\kappa)\,.\label{tetai}\ee Let us introduce
\be \varepsilon(t) = - \frac{\dot{H}}{H^2} = \frac{\dot{\Phi}^2}{2M^2_{Pl}H^2} = \frac{\epsilon_V}{a^6(t)+\frac{\epsilon_V}{3}} \label{epsilon}\ee where we have used the results (\ref{solinfla},\ref{fidotsr},\ref{equ})   from which it is clear that for $\epsilon_V \ll 1$ the slow roll stage begins at $a=1$ when $\varepsilon = \epsilon_V + \mathcal{O}(\epsilon^2_V)$. With $a(t)$ given by (\ref{tila}), it follows that
\be \varepsilon(t) = \frac{3}{\cosh^2[\theta(t)]}\,, \label{vare2} \ee therefore   $0 \leq \varepsilon \leq 3$,    and
\be H(t) = \frac{H_{sr}}{\tanh[\theta(t)]}\,. \label{hubble}\ee

Before we continue with the analysis, it is important to establish the relative variation of the potential between the initial time and the onset of slow roll, assuming that the potential is monotonic and does not feature ``bumps'', this is given by
\be \frac{\Delta V}{V_{sr}} = \Big(\frac{ V'_{sr}}{V_{sr}}\Big)\,\Delta \Phi \label{delVV}\ee where
\be \Delta \Phi = \int^{t_{sr}}_{t_i} \dot{\Phi}(t) dt = \dot{\Phi}_{sr} \int^{t_{sr}}_{t_i} \frac{dt}{a^3(t)}  = \frac{\dot{\Phi}_{sr}}{3H_{sr}}\Big(\frac{3}{\epsilon_V} \Big)^{1/2}\int^{\theta_{sr}}_{\theta_i} \frac{d\theta}{\sinh[\theta]} \label{deltafi}\ee with the result
\be \Delta \Phi = \frac{\dot{\Phi}_{sr}}{3H_{sr}}\Big(\frac{3}{\epsilon_V} \Big)^{1/2} \Bigg\{ \ln \Bigg[\frac{1+f(\kappa)}{1-f(\kappa)} \Bigg]- \ln \Bigg[\frac{1+f(\epsilon_V/3)}{1-f(\epsilon_V/3)} \Bigg] \Bigg\}\,. \label{delfi2}\ee  Using (\ref{slowroll}, \ref{fidotsr}) and (\ref{fofs}) we find
\be \Big|\frac{\Delta V}{V_{sr}}\Big| = 2\sqrt{\frac{\epsilon_V}{3}}  \ln \Bigg[\frac{1+f(\kappa)}{1-f(\kappa)} \Bigg] + \mathcal{O}(\epsilon_V)\,.  \label{delVV2} \ee For large $\kappa$ it follows from (\ref{fofs}) that $f(\kappa) \simeq 1- 1/\sqrt{\kappa}$, hence the logarithm of the term in brackets varies between $  1 -  3$ for $1\leq \kappa \leq 100$, therefore the relative change of the potential during the fast roll stage is $\Delta V/V \simeq \sqrt{\epsilon_V}$ for   $ 1\leq \kappa \lesssim 100$. This result will be used in the next section below to study a systematic
expansion in $\epsilon_V$ to match with the slow roll results.

The acceleration equation written in terms of $\varepsilon(t)$ is given by
\be \frac{\ddot{a}}{a} = H^2(t)(1-\varepsilon(t))\,, \label{acce}\ee so that the inflationary stage begins when $\varepsilon(t) =1$. At the initial time
\be \varepsilon(t_i) = \frac{3\kappa}{1+\kappa} \label{vareini} \ee hence, for $\kappa > 1/2$ the early stage of expansion is deccelerated and inflation begins when $\varepsilon(t_{inf}) =1$.

 It   proves convenient to introduce the variable
 \be x(t) = e^{-\theta(t)/3} = \Big[\frac{\epsilon_V}{12}\Big]^{1/6}\,e^{-H_{sr}t}  \,, \label{xoft} \ee  with
\be x_i\equiv x(t_i) = [f(\kappa)]^{1/3}~~;~~x_{sr}\equiv x(t_{sr}) = [f(\epsilon_V/3)]^{1/3}\, \label{xs}\ee where $f(s)$ is given by eqn. (\ref{fofs}),  and write $a, H,\varepsilon$ in terms of this variable leading to
\be a(x) = \Big[\frac{\epsilon_V}{12} \Big]^{1/6}\,\frac{\big[1-x^6\big]^{1/3}}{x} \,,\label{aofx}\ee
\be H(x)= H_{sr} \frac{\big[1+x^6\big]}{\big[1-x^6\big]} \label{Hofx}\,,\ee
\be \varepsilon(x) = \frac{12\,x^6}{\big[1+x^6\big]^2} \label{epsiofx}\,.\ee

The number of e-folds between the initial time $t_i$ and a given   time $t$ is given by
\be N_e (t;t_i) = \int_{t_i}^{t} H(t')\, dt' = \frac{1}{3}\,\ln\Bigg[    \sqrt{\kappa} \, \frac{(1-x^6(t))}{2x^3(t)}\Bigg]\,,  \label{Ne} \ee with a total number of e-folds between the beginning of the fast roll stage at $t=t_i$ and the onset of slow roll at $t_{sr}$ given by
 \be N_{e}(t_i;t_{sr})=  \frac{1}{6}\,\ln\Big[   \frac{3 {\kappa}}{\epsilon_V} \Big]\,.  \label{Netot} \ee

\begin{figure}[h!]
\begin{center}
\includegraphics[height=3.2in,width=3.0in,keepaspectratio=true]{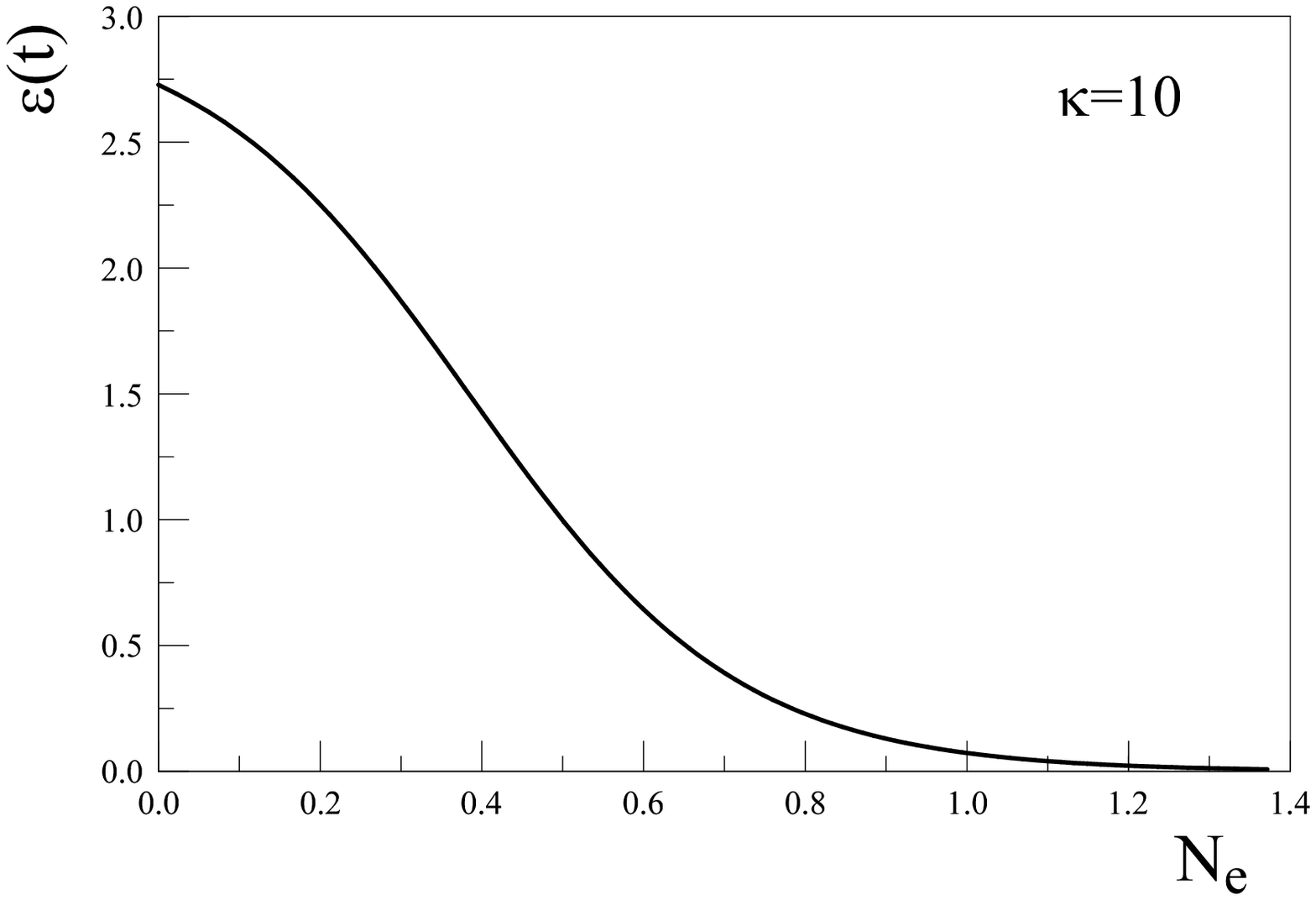}
\includegraphics[height=3.2in,width=3.0in,keepaspectratio=true]{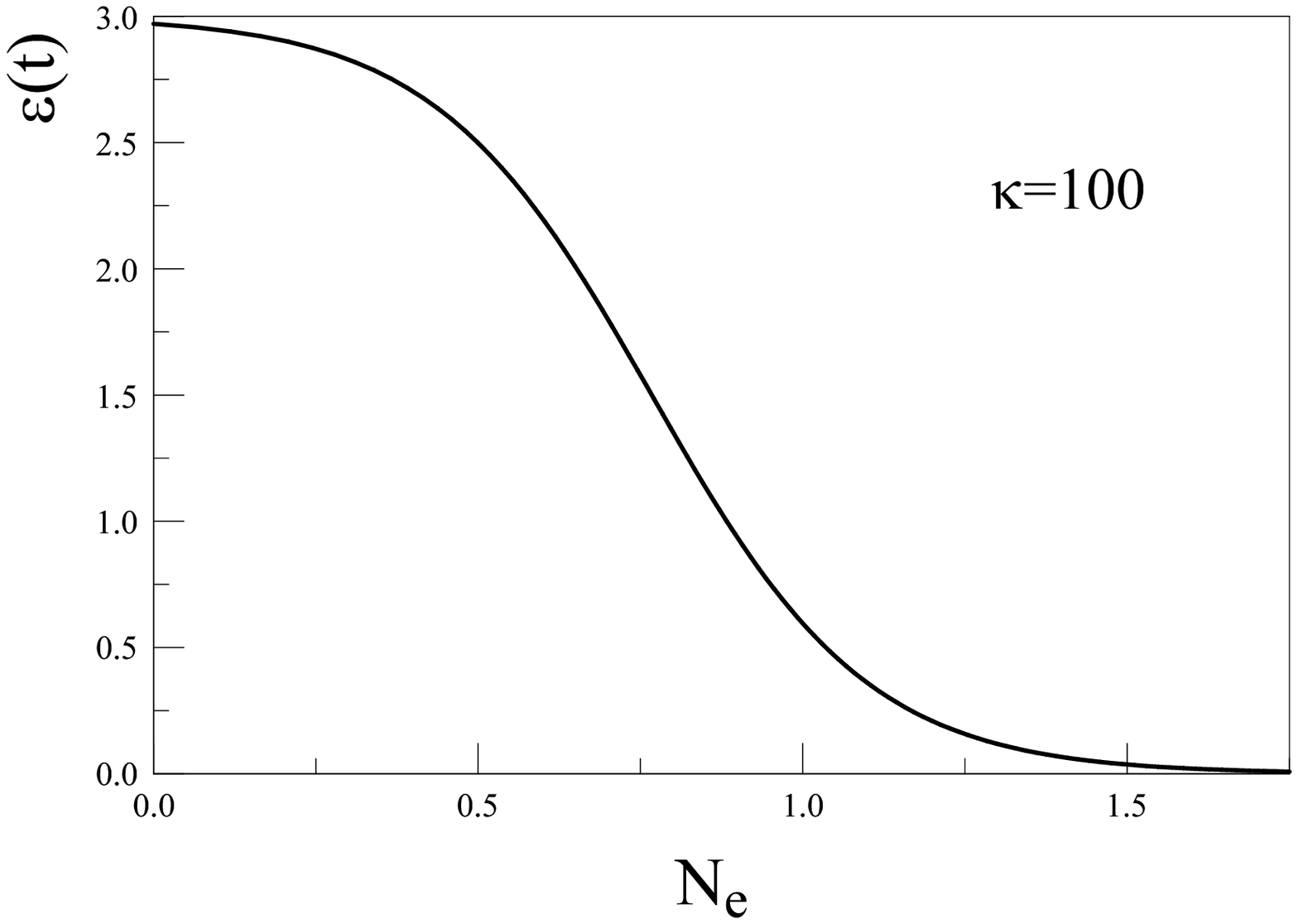}
\includegraphics[height=3.2in,width=3.0in,keepaspectratio=true]{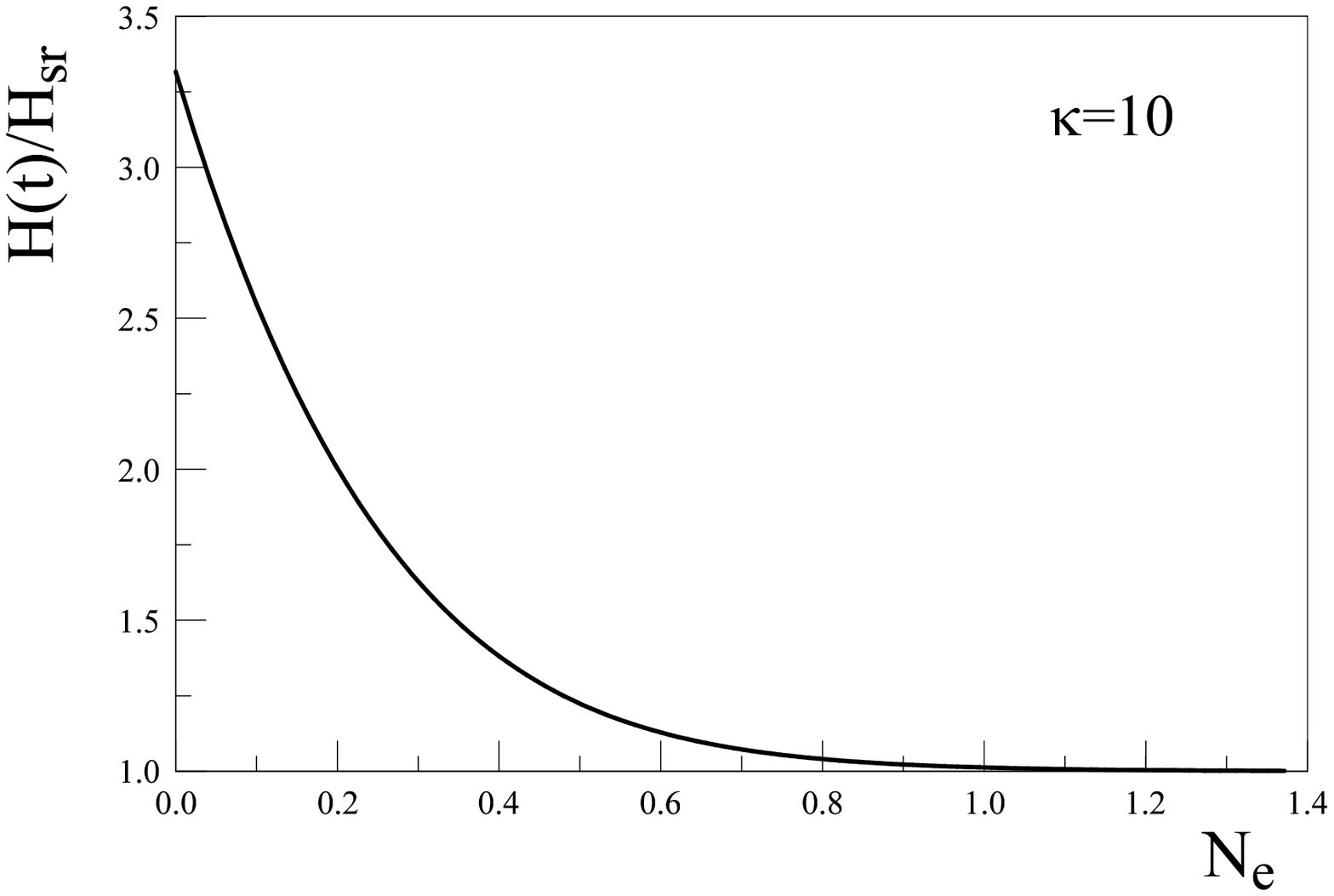}
\includegraphics[height=3.2in,width=3.0in,keepaspectratio=true]{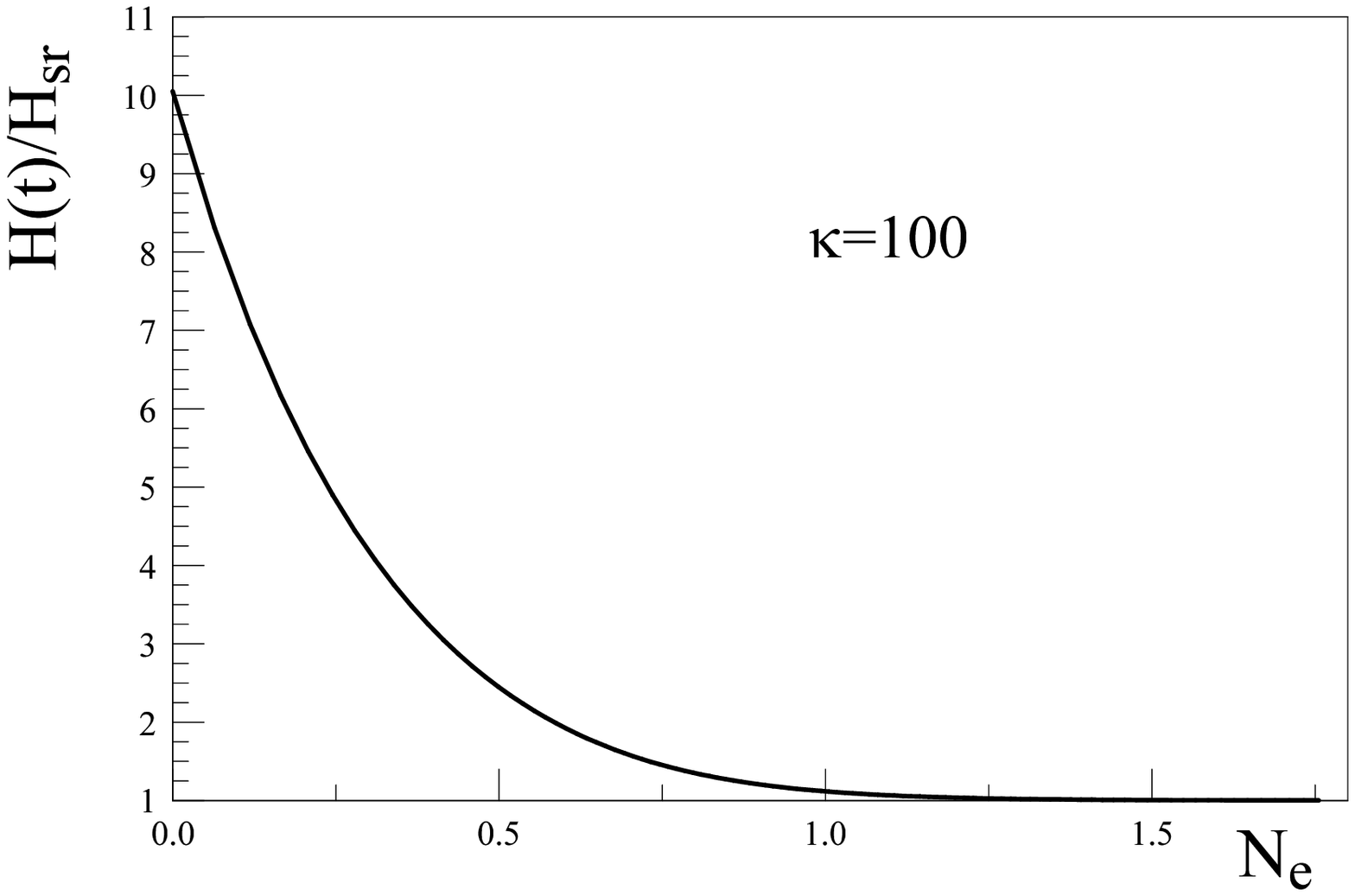}
\caption{$\varepsilon(t)$   and $H(t)/H_{sr}$   as a function of the number of e-folds from the beginning of fast roll, for $\kappa=10;100$ for $\epsilon_V=0.008$. Inflation starts at $N_e \simeq 0.5$, slow roll starts at $N_e \lesssim 1.75$. }
\label{eNe}
\end{center}
\end{figure}

 Fig. (\ref{eNe}) shows $\varepsilon$ as a function of $N_e$ for $\kappa =10;100,\epsilon_V = 0.008$, inflation begins at $N_e \simeq 0.5-0.8$ and slow roll begins at $N_e\simeq 1.37-1.75$. We find that this is the typical behavior for $1 \leq \kappa \leq 100$,  namely for a wide range of fast roll initial conditions, the inflationary stage begins fairly soon $N_{e,inf} \lesssim 1$ and the fast roll stage lasts  $\lesssim 1.7$ e-folds.

\section{Matching to  slow roll:}\label{sec:match}

At the end of the fast roll stage the value of $\dot{\Phi}^2/2V_{sr} \simeq \mathcal{O}(\epsilon_V$), which is of the same order as the slow-roll solution of the equations of motion, and becomes \emph{smaller} than the slow roll solution for $t>t_{sr}$ for which $a(t)>1$. Therefore we must ensure a smooth matching to the slow roll stage. This is accomplished by recognizing that with the fast roll initial conditions there emerges a
 hierarchy of time scales as well as amplitudes for $\dot{\Phi}$: during the fast roll stage the   $\dot{\Phi}$ features a large amplitude $\propto \sqrt{\kappa} \gg 1 $ and varies fast, while in the slow roll stage the amplitude is $\propto \sqrt{\epsilon_V}\ll 1$ and varies slowly. Furthermore, in the previous section   we have taken the potential to be (nearly) constant and recognized that the relative variation during the fast roll stage is $\Delta V/V_{sr} \propto \sqrt{\epsilon_V}$.

  In this section we treat the variation of the potential along with the slow   roll corrections in a consistent perturbative formulation.

   Therefore we write
  \be \Phi(t) = \Phi^{f}(t) + \Phi^{s}(t)~~;~~\Phi^{f}\equiv \Phi^{(0)}, \Phi^{s}\equiv \Phi^{(1)}+ \Phi^{(2)}+\cdots \label{fastslow}\ee where \emph{formally} $\dot{\Phi}^{(0)} \propto (\sqrt{\epsilon_V})^{0}~;~\dot{\Phi}^{(1)} \propto (\sqrt{\epsilon_V}) ~ ; ~ \dot{\Phi}^{(2)} \propto ( {\epsilon_V})\cdots $ with $\dot{\Phi}^{(0)}(t)$
  being the fast roll solution (\ref{fit}) which   is of amplitude $\sqrt{\kappa}$ during most of the fast roll stage. Furthermore, during the fast roll stage we assumed that the potential is nearly constant and equal to the potential during the slow roll stage, namely   $V(\Phi)\simeq V_{sr}$. We now
  relax this assumption by writing  $\Phi= \Phi_{sr}+(\Phi-\Phi_{sr})$ in the argument of the potential $V(\Phi)$ as in eqns. (\ref{delVV}-\ref{delVV2})
  \be V(\Phi)= V_{sr}+ \Delta V(t)~~;~~ \Delta V(t) = V'_{sr}\, \Big( \Delta \Phi(t)\Big) + \frac{1}{2} V''_{sr}\, \Big(\Delta \Phi(t)\Big)^2 + \cdots \,. \label{deltaV}\ee where
  \bea \Delta \Phi(t) & = &   \int^{t}_{\infty} \Big( \dot{\Phi}(t')-\dot{\Phi}_{sr}(t')\Big) dt' = \Delta \Phi^{(0)}(t)+ \Delta \Phi^{(1)}(t)+\cdots \nonumber\\
  \Delta \Phi^{(0)}(t) & = & \int^{t}_{\infty}   \dot{\Phi}^{(0)}(t')  dt' ~~;~~ \Delta \Phi^{(1)}(t)   =   \int^{t}_{\infty} \Big( \dot{\Phi}^{(1)}(t')-\dot{\Phi}_{sr}(t')\Big) dt' \,.\label{delfis} \eea In writing this expression, we have assumed (self-consistently, see below) that   at long time $\dot{\Phi}(t) \rightarrow \dot{\Phi}_{sr}(t)$, hence adjusting an integration constant asymptotically at long time  $ {\Phi}(t) \rightarrow {\Phi}_{sr}(t)$ thereby extending the lower limit of the integral to $t\rightarrow \infty$. This assumption will be justified \textit{a posteriori} from the solution.

  Formally $  V'_{sr}$ is of $\mathcal{O}(\sqrt{\epsilon_V})$, $V''_{sr} \propto \eta_V$ is of $\mathcal{O}( \epsilon_V )$, etc. Therefore
  \be \Delta V = \Delta V^{(1)} + \Delta V^{(2)} +\cdots \,,\label{delVords}\ee  with
  \bea \Delta V^{(1)}  & = &  V'_{sr}  ~ \Delta \Phi^{(0)}(t) \label{delV1} \\
  \Delta V^{(2)}  & = &  V'_{sr} ~\Delta \Phi^{(1)}(t) + \frac{1}{2} V''_{sr}\Big[ \Delta \Phi^{(0)}(t)\Big]^2 \label{delV2} \\
  \vdots & = & \vdots \eea

     Similarly we write
  \be H(t) = \frac{1}{3M^2_{Pl}}\Big[\frac{1}{2}\Big(\dot{\Phi}^{(0)}+\dot{\Phi}^{(1)}+\cdots \Big)^2+V_{sr}+\Delta V]^{1/2}\equiv H^{(0)}+H^{(1)} + H^{(2)}+\cdots \,,\label{hubblefs}\ee where $H^{(0)}$ is the fast roll solution (\ref{frie}) with (\ref{Hsr},\ref{tila}) and
\be H^{(1)} =   \frac{H^2_{sr}}{H^{(0)}}\Big[\frac{\dot{\Phi}^{(0)}\dot{\Phi}^{(1)} }{2 V_{sr}}+\frac{\Delta V^{(1)}}{V_{sr}}\Big]~~;~~ H^{(2)} =   \frac{H^2_{sr}}{H^{(0)}}\Big[\frac{ \big(\dot{\Phi}^{(1)}\big)^2 }{2 V_{sr}}+\frac{\Delta V^{(2)}}{V_{sr}} \Big] ~~;~~\mathrm{etc}\,.\label{H1H2} \ee

With the fast roll solution (\ref{fit},\ref{tila}) we find that
\be \Delta \Phi^{(0)}(t) = \frac{\dot{\Phi}_{sr}}{3H_{sr}}\,\Big(\frac{3}{\epsilon_V} \Big)^{1/2}\,\ln\Big[\frac{1-x^3}{1+x^3} \Big]\label{fidotint}\,,\ee where $x(t)$ is given by eqn. (\ref{xoft}). From this we obtain
\bea \frac{\Delta V^{(1)}}{V_{sr}} & = & 2 \sqrt{\frac{\epsilon_V}{3}}\,\ln\Big[ \frac{1+x^3}{1-x^3}\Big] \,, \label{del1} \\
  \frac{V''_{sr}}{2 V_{sr}}~\Big[ \Delta \Phi^{(0)}(t)\Big]^2 & = & \frac{\eta_V}{3}  \ln^2\Big[ \frac{1+x^3}{1-x^3}\Big]  \,.\label{del22}
\eea

In the equation of motion
\be \ddot{\Phi}+3H\dot{\Phi}= -V'(\Phi)\,  \label{eqmo2} \ee the right hand side is \emph{formally} of order $-V'(\Phi) \propto \sqrt{\epsilon_V} + \cdots $ as can be seen from the definition of the slow roll variable $\epsilon_V$ (\ref{slowroll}). This suggests an expansion in powers of  $\sqrt{\epsilon_V}$ which leads to the following hierarchy of equations
\bea \ddot{\Phi}^{(0)}+3H^{(0)}\dot{\Phi}^{(0)} & = & 0 \label{zeroth}\\
\ddot{\Phi}^{(1)}+3H^{(0)}\dot{\Phi}^{(1)} + 3H^{(1)}\dot{\Phi}^{(0)} & = & -V'_{sr}   \label{first}\\
\vdots & = & \vdots \eea Consistently with the slow roll approximation and to lowest order in slow roll  we neglect $\ddot{\Phi}^{(1)}$ in (\ref{first}) as it can be shown (a posteriori) that $\ddot{\Phi}^{(1)}\propto \big[ 3H^{(0)}\dot{\Phi}^{(1)}\big](\epsilon_V+\eta_V)$, hence higher order in the slow roll expansion.

Inserting the results (\ref{fit}) and (\ref{tila}) for the zeroth order fast roll solution, and the result (\ref{del1}) along with the leading order slow roll relations (\ref{fidotsr}) and the slow roll result
\be \dot{\Phi}_{sr} = -\frac{V'_{sr}}{3 H_{sr}} \label{losr}\ee
into eqn. (\ref{first}) (neglecting $\ddot{\Phi}^{(1)}$), we find
\be  \frac{\dot{\Phi}^{(1)}}{\dot{\Phi}_{sr} } \equiv F[x]=    \frac{T[x]}{2-T^2[x]}\,\Bigg\{1- \frac{2x^3}{3(1+x^6)}\,\ln\Big[ \frac{1+x^3}{1-x^3}\Big]  \Bigg\}~~;~~T[x]=\frac{1-x^6}{1+x^6}\label{funfx}\ee   The function $F[x]$ features the following asymptotic behavior,
\bea && F   \simeq \frac{1}{\sqrt{\kappa}} \Bigg\{1+\frac{1}{6}\ln\big[\frac{\kappa}{4} \big]\Bigg\} ~~\mathrm{for}~~ t \rightarrow t_i ~,~\kappa \gg 1 \label{early} \\
&& F   \simeq  1 +\mathcal{O}(x^6)  ~~\mathrm{for}~~ t \geq  t_{sr}\,. \label{late} \eea

Therefore for $t\rightarrow t_i$ it follows that
\bea \frac{\dot{\Phi}^{(1)}}{\dot{\Phi}^{(0)}} & \simeq & \frac{1}{12 }\,\sqrt{\frac{\epsilon_V}{12}}~\frac{\ln\big[\kappa\big]}{\kappa} \label{dotfiratio}\\
\frac{H^{(1)}}{H^{(0)}} & \simeq & \sqrt{\frac{\epsilon_V}{12}}\,\frac{\ln\big[\kappa\big]}{\kappa}\,.\label{Hratio}\eea
With the results obtained above, it is straightforward to confirm that the second order correction is indeed of $\mathcal{O}(\epsilon_V,\eta_V)$ and further suppressed by a power of $\kappa$ up to logarithmic terms in $\kappa$.

Therefore  up to first order in slow roll
\be \dot{\Phi} = \dot{\Phi}_{sr}\Big[\frac{1}{a^3} + F[x]\Big]\,, \label{uptofi}\ee
\be H^2 = H^2_{sr}\Bigg\{1+\frac{\epsilon_V}{3}\Big[\frac{1}{a^3}+F[x] \Big]^2\Bigg\} \,.\label{h2u}\ee
The second order contribution $\Delta V^{(2)}/V_{sr}$ can be found by carrying out the integral in the second term in (\ref{delfis}), this is achieved more efficiently by passing to the variable $x$ and expanding the function $x$ in a series in $x^3$ and integrating term by term. The result reveals that this correction is $\mathcal{O}(\epsilon_V,\eta_V)$ and suppressed by a power of $\kappa$ as $t\rightarrow t_i$ and is also subleading for $t\geq t_{sr}$.

The result (\ref{funfx}) clearly shows that for $t\gtrsim t_{sr}$
\be \dot{\Phi}^{(1)}-\dot{\Phi}_{sr} = \dot{\Phi}_{sr}\Big[F[x]-1\Big] \simeq \epsilon_V\,e^{-6H_{sr}t} \label{asyfidots}\ee (see eqn. (\ref{xoft})) so that adjusting the integration constant in eqn. (\ref{asyfidots}) so that $\Phi^{(1)}\rightarrow \Phi_{sr}$ as $t\rightarrow \infty$ justifies  the assumption that asymptotically $\Phi -\Phi_{sr} \rightarrow 0$ as $t\rightarrow \infty$ thus validating  the expressions (\ref{delfis}).

For $t \rightarrow t_i$ these expressions reduce to the fast roll results of the previous section; however, for $t>t_{sr}$ ($x^3 \lesssim (\epsilon_V/12)^{1/2}$) we find
\bea \dot{\Phi} & \rightarrow &  \dot{\Phi}_{sr} + \mathcal{O}\Big(\epsilon_V\,e^{-6H_{sr}t}\Big)\label{dotfitgreat}\\
H^2 & \rightarrow & H^2_{sr}\Big[1+\frac{\epsilon_V}{3}\Big]\,. \label{Hfitgreat}\eea

The correction to the scale factor is obtained by proposing a solution of the form
\be a(t) = a^{(0)}(t)~ a^{(s)}(t)  \label{scalefactor}\ee where
\be \frac{\dot{a}^{(0)}}{a^{(0)}} = H^{(0)}~,~ \frac{\dot{a}^{(s)}}{a^{(s)}} = H^{(1)}+H^{(2)}+\cdots~  \label{ahs}\ee It is straightforward to find that asymptotically for $t\gg t_{sr}$
\be a^{(0)}(t) = e^{H_{sr} t}~,~a^{(s)}(t) \rightarrow   \Big[ a^{(0)}(t) \Big]^{\epsilon_V/6} \label{asyscafa}\ee where a detailed calculation shows that the terms proportional to $\Delta V$ are subleading for $t\gg t_{sr}$. Therefore
 the improved fast roll solution (\ref{h2u}) yields the \emph{correct} long time behavior of the scale factor to leading order in slow roll; namely, for $t > t_{sr}$, the dynamics enters a near de Sitter stage
\be \ln[a(t)] \rightarrow H_{sr}\Big[ 1 +\frac{\epsilon_V}{6}
\Big]~t \,.\label{asyaoft}\ee

It is now clear from the solution (\ref{uptofi})  that for $t \ll t_{sr}$ the fast roll, zeroth order solution ($\propto 1/a^3$) dominates but at $t\simeq t_{sr}$  ($a(t_{sr}) =1), F[x_{sr}]\simeq 1$ and $\dot{\Phi} \approx 2 \dot{\Phi}_{sr}$. Therefore, at $t_{sr}$, the solution is of order $\sqrt{\epsilon_V}$ but off by a factor $2$ from the correct solution, resulting in an error of $\mathcal{O}(\epsilon_V)$. In order to match to the correct slow roll solution the evolution must be continued past $t_{sr}$ to a time $t_{m}$ at which the first order correction dominates. This ``matching time'', $t_m$, is determined by the error incurred in keeping the zeroth order term in the full solution. For example, requiring that the error be $\simeq \epsilon_V\, \sqrt{\epsilon_V}$ fixes $t_m$ so that
\be a^3(t_m) \simeq \frac{1}{\sqrt{\epsilon_V}} \Rightarrow x^3_m \simeq  \Big( \frac{\epsilon^2_V}{12}\Big)^{1/2} \label{match}\ee hence, at the ``matching time'', we find that
\be \dot{\Phi}(t_m) = \dot{\Phi}_{sr} \Bigg[1+ \mathcal{O}(\sqrt{\epsilon_V})\Bigg]\,. \label{matchfidot} \ee The number of e-folds between the time $t_{sr}$, at which $\dot{\Phi}^{(0)} \simeq \dot{\Phi}_{sr}$, and the matching time $t_m$ is
\be N_e(t_{sr};t_{m}) = -\frac{1}{6}\ln[\epsilon_V] \simeq 0.8 \label{efoldmatch}\ee where the numerical result applies for $\epsilon_V=0.008$. Therefore for $\kappa \lesssim 100, \epsilon_V = 0.008$, the total number of e-folds between the initial and the matching time is $N_e \simeq 2.5 $.

With the improved solution (\ref{uptofi}), it follows that the variable $\varepsilon(t)$ defined by eqn. (\ref{epsilon}) is given by
\be \varepsilon(t) =     \frac{\epsilon_V~\Big[\frac{1}{a^3} + F[x]\Big]}{1+\frac{\epsilon_V}{3}\Big[\frac{1}{a^3} + F[x]\Big]}\,. \label{impepsi}\ee This quantity is a better indicator of the transition to slow roll, it features the following limits
\bea \varepsilon(t) & \simeq &  3 ~~ \mathrm{for}~~ t\simeq t_i, \kappa \gg 1\nonumber \\
\varepsilon(t) & \simeq &  2 \epsilon_V ~~ \mathrm{for}~~ t\simeq t_{sr} \\ \nonumber
\varepsilon(t) & \simeq &  \epsilon_V  ~~ \mathrm{for}~~ t > t_{sr}\,. \label{limsvarepsi}\eea with corrections of $\mathcal{O}( \epsilon^{3/2}_V) $ at the matching time $t_m$.

 Conformal time $\eta(t)$ defined to vanish as $t\rightarrow \infty$  is given by
\bea \eta(t)  & = &  \int^{t}_{\infty} \frac{dt'}{a(t')} = \int^{a(t)}_{\infty} \frac{da}{a^2\,H(a)} \nonumber \\ & = & -\frac{1}{a(t)H(t)} +\int^{t}_{\infty} \varepsilon(t')\frac{dt'}{a(t')} \label{etadef}\eea where we integrated by parts and used the definition of $\varepsilon$ given by eqn. (\ref{epsilon}). Adding and subtracting $\epsilon_V$  we find

\be \eta(t)  = -\frac{1}{a(t)H(t)(1-\epsilon_V)}+ \frac{\epsilon_V}{ \big(1-\epsilon_V\big)}\int_{\infty
}^{t}\,\Big[\frac{\varepsilon(t')}{\varepsilon_V}-1\Big] \frac{dt'}{a(t')}    \,, \label{etadef2} \ee

The argument of the integrand in the second term in (\ref{etadef2}) vanishes to leading order in $\epsilon_V,\eta_V$  in the slow roll phase (when $t>t_{sr}$). Therefore, during slow roll, $\eta = -1/aH(1-\epsilon_V)$.

\vspace{2mm}

\textbf{Discussion:}

The study in this section describes a systematic procedure to  obtain a solution that is valid during the fast roll stage and that matches smoothly to the slow roll stage   to any desired order in $\epsilon_V,\eta_V$ \emph{independently of the potential} while under the assumption that the inflationary potential is monotonic and can be  described  by a derivative expansion characterized by slow roll parameters. The leading order solution is the fast roll solution (obtained in the previous section) and the above analysis shows that continuing this solution for time larger than $t_{sr}$ incurs errors of order $  \epsilon_V$ in the variable $\varepsilon$: at $t=t_{sr}$ the zeroth-order and the improved solution differ by $\epsilon_V$ which in turn leads to corrections $ \leq \epsilon_V^2$ in the conformal time $\eta$.

This analysis shows that the leading order corrections to the inflationary power spectra from a fast roll stage can be obtained by keeping only the fast-roll solution and integrating up to $t\simeq t_{sr}$, at which point it matches to slow roll. Clearly keeping only the zeroth-order solution rather than the improved solution incurs errors of $\mathcal{O}(\epsilon_V)$, which can (with numerical effort) be   systematically improved upon by considering the corrections and improvements described in this section.

Having quantified the error incurred in keeping only the fast roll solution, we now proceed to obtain the corrections to the power spectra of scalar and tensor perturbations \emph{to leading order in the expansion in slow roll parameters}, namely keeping only the fast roll solution.

\section{Fast roll corrections to power spectra:}

The analysis above clearly indicates that for a wide range of initial conditions dominated by the kinetic term of the inflaton potential, a fast roll stage merges with the slow roll stage within $2-3$ e-folds.
Having quantified above the error incurred in keeping only the fast roll solution, we now proceed to obtain the corrections to the power spectra of scalar and tensor perturbations \emph{to leading order in the expansion in slow roll parameters}, namely keeping only the fast roll solution. The results will yield the main corrections to the power spectra from the fast roll stage, with potential corrections of $\mathcal{O}(\epsilon_V)$ from the matching of scales. If the main features of the results obtained in leading order are supported observationally, this would justify a more thorough study that includes these corrections by implementing the systematic approach described in the previous section. Such a program would necessarily imply a larger numerical effort and would be justified if observational data suggest the presence of the main effects.

The observational constraint of nearly scale invariance suggests that wavelengths corresponding to observable quantities today crossed the Hubble radius during the slow roll era of inflation.
Therefore our goal is to analyze the impact of the pre-slow roll dynamics upon perturbations with physical wavelengths that crossed the Hubble radius \emph{after} the beginning of slow roll.
As discussed in refs.\cite{boyan3,hectordestri} and more recently in ref.\cite{lellodS} the fast-roll stage prior to slow roll modifies the initial conditions on the mode functions from the usual Bunch-Davies case.
The rapid dynamical evolution of the inflaton during the fast roll stage induces a correction to the potential in the equations of motion for the mode functions of curvature and tensor perturbations, which we now analyze in detail.

The gauge invariant curvature perturbation of the comoving
hypersurfaces is given in terms of the Newtonian potential ($\psi(\vec{x},t)$)
 and the inflaton fluctuation  ($\delta \phi(\vec{x},t)$) by\cite{mukh,kolb,riotto2,baumann,giov}
\be \label{curvature} \mathcal{R}= -\psi-\frac{H}{\dot{\Phi} } \;
\delta \phi  \; . \ee where $ \dot \Phi  $ stands for the derivative of the
inflaton field $ \Phi  $ with respect to the cosmic time $ t $.

It is convenient to introduce the gauge invariant potential
\cite{mukh,kolb,riotto2,baumann,giov}, \be u(\vx,t)=-z \;  \mathcal{R}(\vx,t) \label{u}
\; , \ee where \be \label{za} z= a(t) \; \frac{\dot{\Phi} }{H} \;  .
\ee The gauge invariant field $
u(\vx,t) $  is quantized by   expanding in terms of conformal time mode functions and
creation and annihilation operators as follows\cite{mukh,kolb,riotto2,baumann,giov}
\be \label{curvau} u(\vec{x},\eta)= \frac{1}{\sqrt{V}} \sum_{\vec{k}}\Big[
\alpha_\mathcal{R}(k)\,S_\mathcal{R}(k;\eta)\,e^{i\vec{k}\cdot\vec{x}}+
\alpha^\dagger_\mathcal{R}(k)\,S^*_\mathcal{R}(k;\eta)\,e^{-i\vec{k}\cdot\vec{x}}\Big] \,. \ee The operators $ \alpha_\mathcal{R}(k),\alpha^\dagger_\mathcal{R}(k)
$ obey canonical commutation relations and the mode functions are solutions of the  equation \be
\Bigg[\frac{d^2}{d\eta^2}+k^2-   \frac{z''}{z}
\Bigg]S_\mathcal{R}(k;\eta) =0 \,. \label{Scureq} \ee

 Tensor perturbations (gravitational waves) correspond to
minimally coupled  massless fields with two physical transverse polarizations, the
quantum fields are written as \cite{mukh,kolb,riotto2,baumann,giov} \be
h_{ij}(\vec{x},\eta) = \frac{2}{ C(\eta) \, M_{Pl}      }\sum_{\vec{k}}
\sum_{\lambda=\times,+} \epsilon_{ij}(\lambda,\vec{k})
\left[\alpha_{\lambda,\vec{k}} \; \, S_T(k;\eta)\,e^{i\vec{k}\cdot\vec{x}}+
\alpha^\dagger_{\lambda,\vec{k}} \; \, S^*_T(k;\eta)\,e^{-i\vec{k}\cdot\vec{x}} \right] \; ,
\label{tens} \ee   where $ \lambda $ labels the two standard
transverse and traceless polarizations $ \times $ and $ + $. The
operators $ \alpha_{\lambda,\vec{k}}, \;
\alpha^\dagger_{\lambda,\vec{k}} $ obey the usual canonical
commutation relations, and $ \epsilon_{ij}(\lambda,\vec{k}) $ are
the two independent traceless-transverse tensors constructed from
the two independent polarization vectors transverse to $
\hat{\bf{k}} $, chosen to be real and normalized such that $
\epsilon^i_j(\lambda,\vec{k})\, \;
\epsilon^j_k(\lambda',\vec{k})=\delta^i_k
 \; \delta_{\lambda,\lambda'} $.

The mode functions $ S_T(k;\eta) $ obey the differential equation of a massless minimally coupled scalar field, namely
\be\label{Sten} \left[ \frac{d^2}{d\eta^2} + k^2-
\frac{C''(\eta)}{C(\eta)}\right]S_{T}(k;\eta) = 0 \,.  \ee In both these cases
 the mode functions obey an equation of
the form,
\be
\left[ \frac{d^2}{d\eta^2}+k^2-W_\alpha(\eta) \right]S_\alpha(k;\eta) =0~~;~~\alpha=R,T\,. \label{modi} \ee
 This is  a
Schr\"odinger equation with $ \eta $ playing the role of coordinate, $ k^2 $ the
energy and $ W(\eta) $ a potential that depends on the coordinate $
\eta $. In the cases under consideration \be W_\alpha(\eta)  = \Bigg\{
\begin{array}{l}
z''/z  ~~\mathrm{for~curvature~perturbations}          \\
C''/ C  ~~\mathrm{for~tensor~perturbations} \\
\end{array} \,.\label{defiW}
\ee During slow roll inflation the potential $W_\alpha(\eta)$ becomes
\be\label{defV}
W_\alpha(\eta)=
\frac{\nu^2_\alpha-\frac14}{\eta^2} \; ,
\ee where to leading order in slow roll parameters
\be \nu_\alpha  = \frac{3}{2}+ \Bigg\{
\begin{array}{l}
 3\epsilon_V -\eta_V ~~\mathrm{for~curvature~perturbations}          \\
 \epsilon_V   ~~\mathrm{for~tensor~perturbations} \\
\end{array} \,.\label{defW}
\ee

The full dynamical evolution of the inflaton during the fast roll stage leads to a modification of the mode equations (\ref{modi}) over terms of a potential $ {V}_\alpha(\eta)$ that is localized in $\eta$ in a narrow range prior to the slow roll phase\cite{boyan3,hectordestri,reviunos}. Specifically, in the mode equations (\ref{modi}), $W(\eta)$ is modified as
\be W_\alpha(\eta) =  {V}_\alpha(\eta) + \frac{\nu^2_\alpha-1/4}{\eta^2} ~~;~~ V_\alpha(\eta) = \Bigg\{\begin{array}{l}
                \neq 0 ~\mathrm{for}~~ \eta_i < \eta < {  \eta_{sr}} \\
                0 ~\mathrm{for} ~ {  \eta_{sr}} < \eta \,,  \\
              \end{array}  \label{Vpot} \ee where $\nu_\alpha$ is given by (\ref{defW}) for curvature and tensor perturbations.

              For curvature perturbations  we find
\be W_R(\eta) = \frac{z''}{z} = \frac{a^2}{z} [\ddot{z}+H\dot{z}] =     2 a^2 H^2 \Bigg[1-\frac{7}{2}\varepsilon+\varepsilon^2+(3-\varepsilon)\Big[2\sqrt{\varepsilon\epsilon_V}-
\frac{\eta_V}{2} \Big] \Bigg] \label{curvapot} \ee Therefore, to leading order in slow roll, the potential for curvature perturbations is given by
\be  {V}_R(\eta) = W_R(\eta)-\frac{2}{\eta^2}\Big[1+\frac{9}{2}\epsilon_V-\frac{3}{2}\eta_V\Big]\,. \label{Vcurva}\ee

For tensor perturbations
 \be W_T(\eta) = \frac{C''}{C} = a [\ddot{a}+H\dot{a}] =     2 a^2 H^2 \Big[1- \frac{\varepsilon}{2}\Big]\,,  \label{tensorpot} \ee and the potential for tensor perturbations is given by
\be  {V}_T(\eta) = W_T(\eta)-\frac{2}{\eta^2}\Big[1+\frac{3\epsilon_V}{2}\Big]\,, \label{vtensor}\ee

The potentials as a function of $\eta$ are found parametrically in terms of the dimensionless variable $x$, given by (\ref{xoft}), by writing $a,H,\varepsilon, \eta$ all as functions of $x$.

 From the result of the previous sections, it is clear that as $t\rightarrow t_{sr}$, $\varepsilon \rightarrow \epsilon_V$ and $a^2 H^2 \rightarrow 1/\eta^2$, higher order corrections in $\epsilon_V$ arise from the matching to the slow roll stage thereby yielding higher order corrections in $\epsilon_V,\eta_V$ to the potentials $V_{R,T}$.

 Therefore we focus on obtaining the leading order effects from the fast-roll stage by considering solely the fast-roll solution   given by eqns.(\ref{aofx}-\ref{epsiofx}) along with replacing the fast roll solution (\ref{epsilon}) into the expression for $\eta$ (\ref{etadef2}). This yields
 \bea \eta (x) & = &  -  \frac{1}{H_{sr}\,\big(1-\epsilon_V\big)} \, \Big( \frac{12}{\epsilon_V}\Big)^{1/6} \Bigg\{ \frac{x(1-x^6)^{2/3}}{(1+x^6)} \nonumber \\ &+ & \epsilon_V \,\int^{x}_{x_{sr}} \frac{dy}{[1-y^6]^{1/3}}\,
\Bigg[ \frac{12}{\epsilon_V} \frac{y^6}{(1+y^6)^2}-1\Bigg] \Bigg\}\,. \label{etafx}\eea
where the lower limit in the integral ensures the matching to the slow roll result at $t_{sr}$.  The potentials are now obtained to leading order by replacing the expressions (\ref{aofx}-\ref{epsiofx},\ref{etafx}) into (\ref{curvapot},\ref{Vcurva}) and in (\ref{tensorpot},\ref{vtensor}). As discussed in the previous sections, considering the lowest order solutions captures the full fast roll stage and yields an error $\simeq \mathcal{O}(\epsilon_V)$ for $t>t_{sr}$ during the slow roll stage.

\begin{figure}[h!]
\includegraphics[height=3.2in,width=3.2in,keepaspectratio=true]{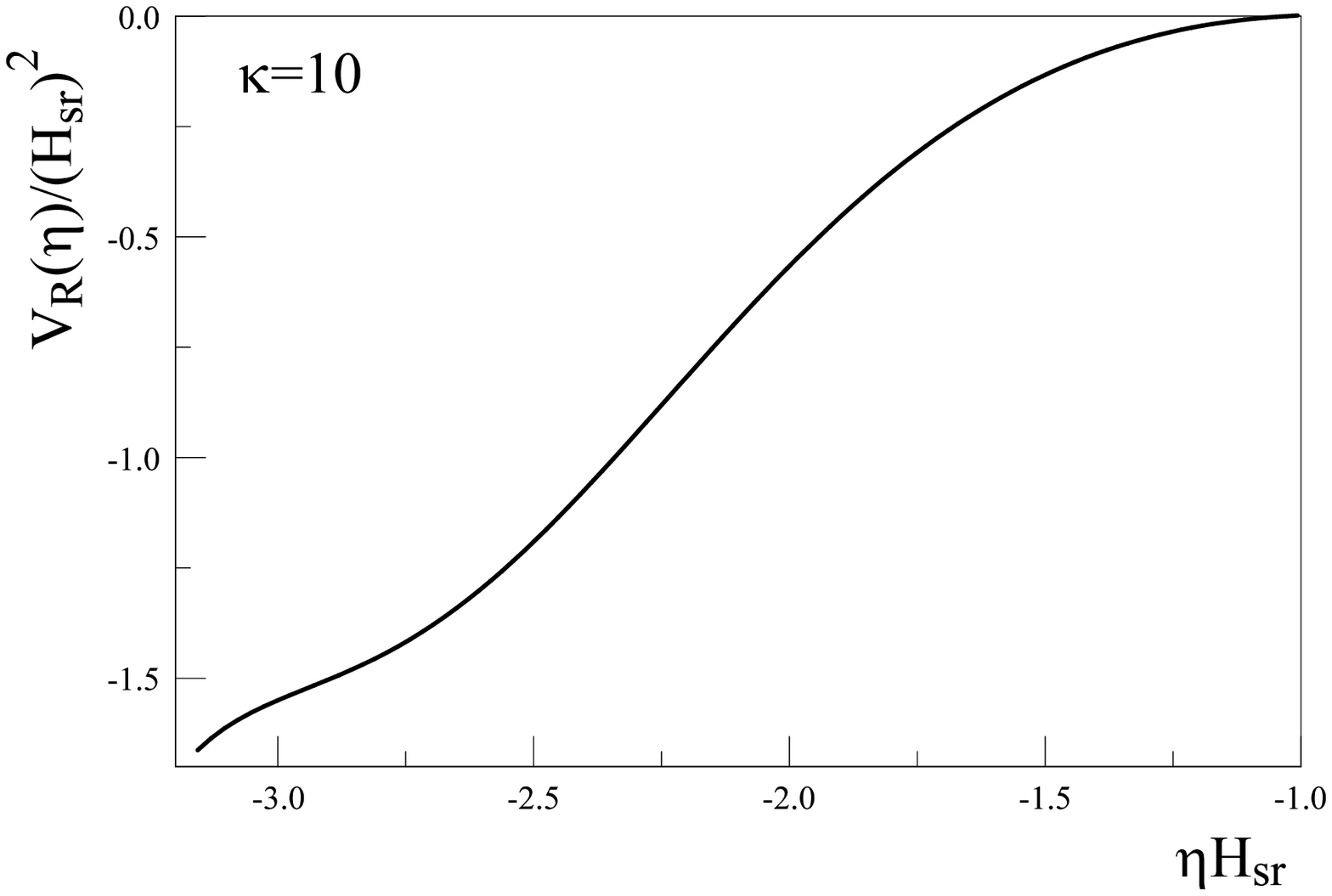}
\includegraphics[height=3.2in,width=3.2in,keepaspectratio=true]{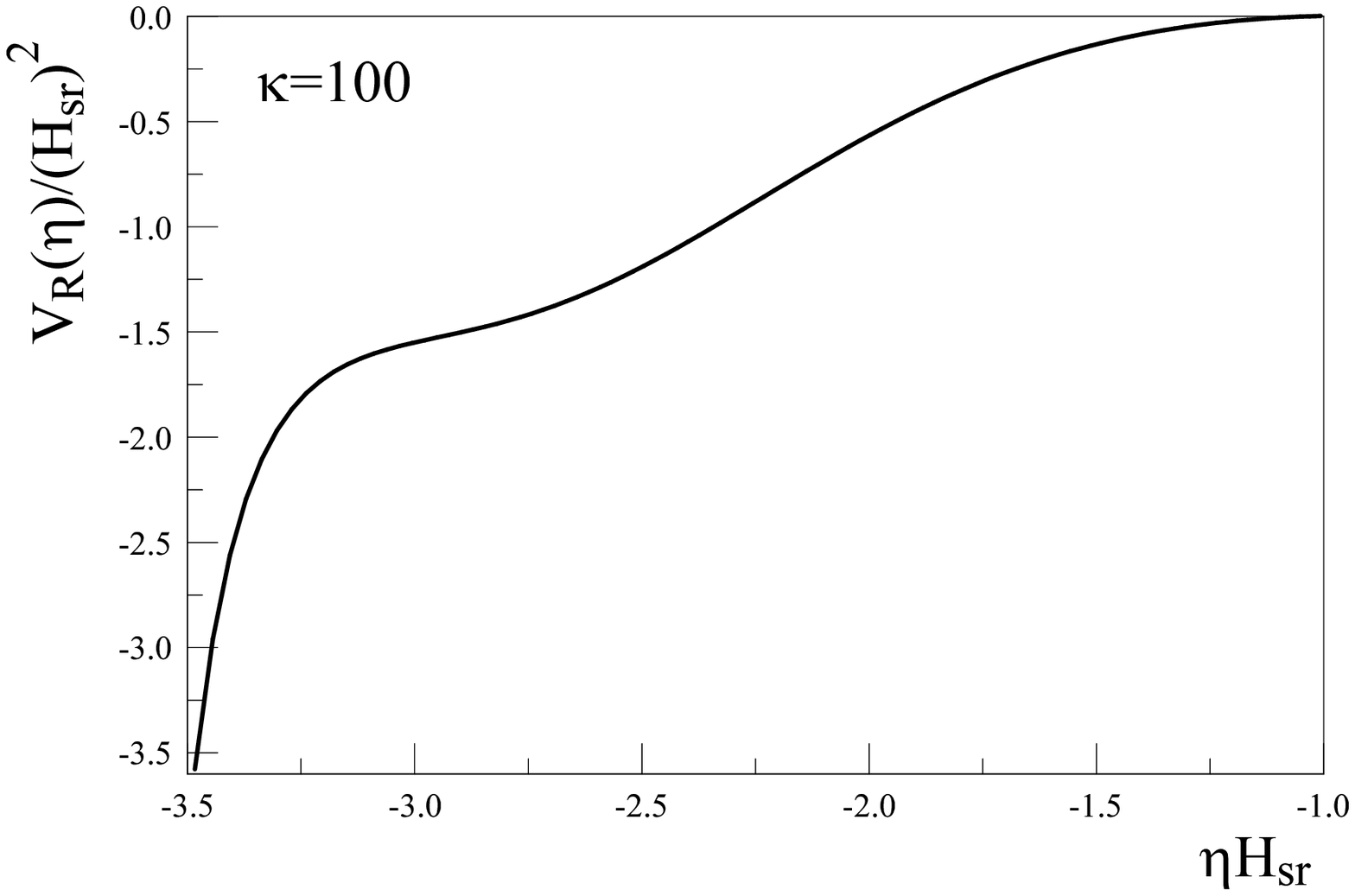}
\caption{Potentials for curvature perturbations $ {V}_R(\eta)$     as a function of $\eta$ from the beginning of fast roll, for $\kappa=10;100~;~\epsilon_V=0.008~;~\eta_V=-0.010$.  }
\label{fig:curvapotentials}
\end{figure}

\begin{figure}[h!]
\includegraphics[height=3.2in,width=3.2in,keepaspectratio=true]{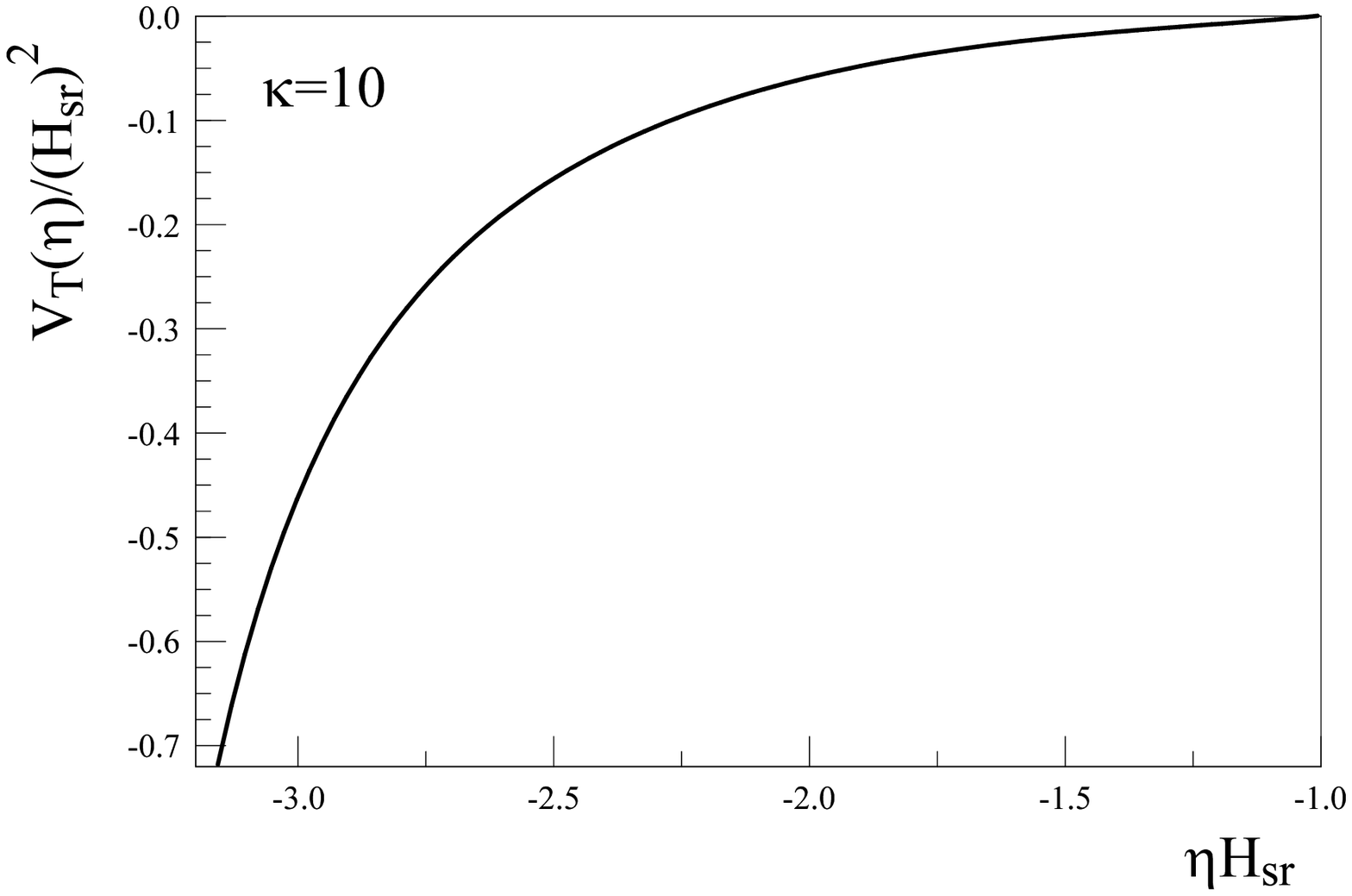}
\includegraphics[height=3.2in,width=3.2in,keepaspectratio=true]{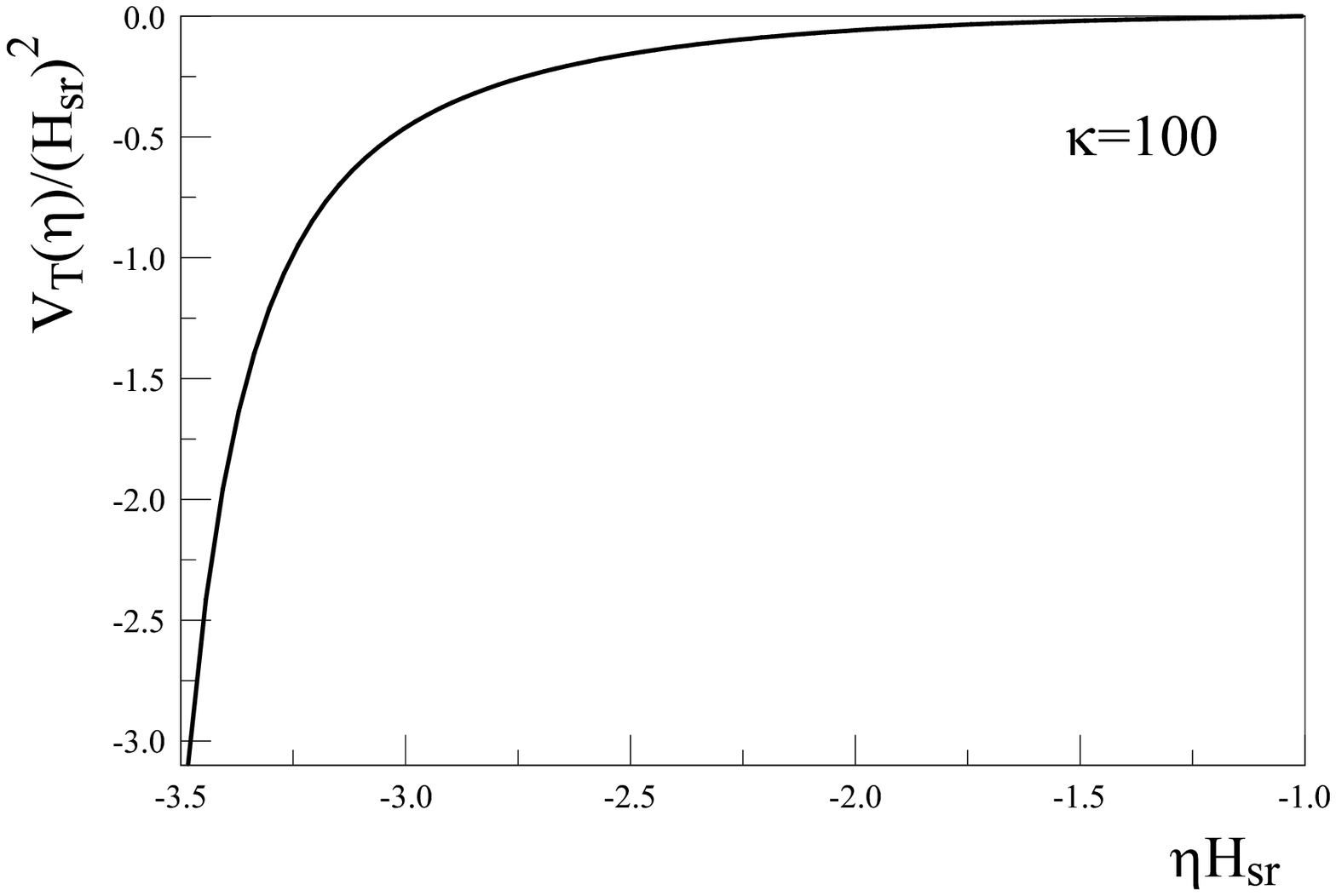}
\caption{Potentials for tensor perturbations $ {V}_T(\eta)$     as a function of $\eta$ from the beginning of fast roll, for $\kappa=10;100~;~\epsilon_V=0.008~;~\eta_V=-0.010$.  }
\label{fig:tensorpotentials}
\end{figure}

The potentials $V_R(\eta)~;~V_T(\eta)$ are shown in figs. (\ref{fig:curvapotentials},\ref{fig:tensorpotentials}) for $\kappa=10;100$ for $\epsilon_V =0.008~;~\eta_V=-0.010$.

These potentials are qualitatively similar to those found for a specific choice of the inflaton potential and initial conditions in the second reference in\cite{boyan3}. Two important aspects explain most of the quantitative discrepancy between our results and those of this reference: i) the potential scales $\propto a^2$, therefore by normalizing the scale factor to unity at the beginning of slow roll, our potential features an overall scale with respect to that in ref.\cite{boyan3}, ii) $\eta$ scales $\propto 1/a$ therefore there is also an overall scaling in the definition of conformal time, our normalization is more convenient to analyze the transition to slow roll and assess when wavevector cross the Hubble radius during slow roll inflation. Furthermore the particular choice of the potential in this reference also modifies the values of $\epsilon_V,\eta_V;\kappa$.  Accounting for the different normalizations (scale of $a$ affecting the definition of the potential and $\eta$), the similarity of the potentials is both reassuring and expected because the potential is determined by the fast roll stage which is dominated by the fast evolution of the inflaton field and is rather insensitive to the potential as long as the potential is sufficiently flat to be consistent with slow roll. Thus our results are robust and to leading order in slow roll variables only depend  on $\kappa$, $\epsilon_V$ and $\eta_V$ \emph{regardless of the specific form of the inflationary potential}.

During the slow roll stage the solution of the mode equations (\ref{modi}) with ${V}_\alpha(\eta)=0$, namely with $W_\alpha(\eta)$ given by (\ref{defV}), are
 \be S_{\alpha}(k;\eta) = A_{k,\alpha} \,g_{\nu_\alpha}(k;\eta) + B_{k,\alpha}\,g^*_{\nu_\alpha}(k;\eta) ~~;~~\alpha=T,R\,. \label{Ssr}\ee where, up to an overall phase,
\be   g_{\nu}(k,\eta) =   \sqrt{\frac{-\pi \eta}{4}} \, H^{(1)}_{\nu} (-k \eta)\, \label{BDmodes} \ee are the solutions with Bunch-Davies initial conditions, where $\nu_\alpha$ is given by (\ref{defW}) for curvature and tensor perturbations.

The power spectra for  curvature   $(\mathcal{R})$ and tensor (gravitational waves) (T) perturbations are respectively
\be  \mathcal{P}_{\mathcal{R}}(k) = \frac{k^3}{2\pi^2} \Bigg|\frac{S_{\mathcal{R}}(k;\eta)}{z(\eta)}\Bigg|^2~~;~~\mathcal{P}_{T}(k) = \frac{4\,k^3}{ \pi^2\,M^2_{pl}} \, \Bigg|\frac{S_T(k;\eta)}{C(\eta)}\Bigg|^2 \,. \label{powerspectra}\ee We assume that the modes of cosmological relevance today crossed the Hubble radius during the slow roll inflationary stage, therefore evaluating these power spectra a few e-folds after horizon crossing $-k\eta \ll 1$ during slow roll,  it follows that  the  general solution (\ref{Ssr}) is given by
\be S_\alpha(k;\eta)  =   -i\,\sqrt{\frac{-\pi \eta}{4}}\,\frac{\Gamma(\nu_\alpha)}{\pi}\,\Big(-\frac{k\eta}{2}\Big)^{-\nu_\alpha}  \big[A_{k,\alpha}-B_{k,\alpha} \big]~~;~~-k\eta \ll 1 \; , \label{BDmodeslargeeta} \ee therefore
the power spectra become
\be \mathcal{P}_{\alpha}(k) =   \mathcal{P}^{BD}_{\alpha}(k)\,\mathcal{T}_\alpha(k)~~;~~ \alpha= \mathcal{R},T \,, \label{transfer} \ee where $\mathcal{P}^{BD}_{\alpha}(k)$ are the power spectra for Bunch-Davies modes $g_\nu(k;\eta)$, namely for  $A_k =1;B_k=0$, and
\be \mathcal{T}_{\alpha}(k) = \big|A_{k,\alpha} - B_{k,\alpha}\big|^2 \label{tfun}\ee is a transfer function that encodes the non-Bunch-Davies initial conditions for the respective perturbations.

 With $W_\alpha(\eta)$ given by (\ref{defV},\ref{defW}) during slow roll, we find
\be z(\eta) = z_0\Big(\frac{-\eta}{-\eta_0}\Big)^{\frac{1}{2}-\nu_{\mathcal{R}}} ~~;~~z_0 = \Bigg[\frac{a\dot{\Phi}}{H}\Bigg]_{\eta_0}\,,\label{zetaeta}\ee and
\be C(\eta) = C_0\Big(\frac{-\eta}{-\eta_0}\Big)^{\frac{1}{2}-\nu_{T}} ~~;~~C_0 = \Bigg[\frac{1}{-\eta\,H}\Bigg]_{\eta_0}\,,\label{Ceta}\ee where $C_0$ given in eqn. (\ref{Ceta}) is to leading order in slow roll and $\eta_0$ is an arbitrary scale. Therefore, to leading order in slow roll, we find that
\be \mathcal{P}_{\mathcal{R}}(k) = \frac{1}{4\pi^2\,z^2_0\,\eta^2_0} \Big(-k\eta_0\Big)^{n_s-1} \, \mathcal{T}_{\mathcal{R}}(k)~~;~~n_s-1=-6\epsilon_V+2\eta_V\,. \label{powerR1}\ee
Using the value $z_0$ in eqn.(\ref{zetaeta}), along with the slow roll relation $\dot{\Phi}^2/H^2 = 2 M^2_{Pl}\epsilon_V$ and defining $-\eta_0 \equiv 1/k_0$ as a ``pivot'' scale, we finally find to leading order in slow roll
\be \mathcal{P}_{\mathcal{R}}(k) = \frac{H^2}{8\pi^2\,M^2_{Pl}\epsilon_V}\Big|_{ \eta_0=-1/k_0} \Big(\frac{k}{k_0}\Big)^{n_s-1}\, \mathcal{T}_{\mathcal{R}}(k)\,. \label{PRfini}\ee Therefore, choosing $k$ as the ``pivot'' scale $k_0$, gives us
\be \mathcal{P}_{\mathcal{R}}(k_0) = \frac{H^2}{8\pi^2\,M^2_{Pl}\epsilon_V}\Big|_{-\eta_0=1/k_0} \, \mathcal{T}_{\mathcal{R}}(k_0)\,. \label{PRfini2}\ee Carrying out similar steps for tensor perturbations
we find
\be \mathcal{P}_{T}(k) = \frac{2}{ \pi^2\,M^2_{Pl}C^2_0\,\eta^2_0 } \Big(-{k}{\eta_0}\Big)^{n_T}\, \mathcal{T}_{T}(k)~~;~~n_T = -2\epsilon_V \,. \label{PTfini}\ee Using $C^2_0 \eta^2_0 = 1/H^2$ from (\ref{Ceta})  (to leading order in slow roll), leads to
\be \mathcal{P}_{T}(k) = \frac{2 H^2}{ \pi^2\,M^2_{Pl}}\Big|_{\eta_0 = -1/k_0}\, \Big(\frac{k}{k_0}\Big)^{n_T}\, \mathcal{T}_{T}(k)\,.  \label{Ptensorf}\ee Therefore the tensor to scalar ratio,  is given by
\be r(k ) = \frac{ \mathcal{P}_{T}(k)}{\mathcal{P}_{\mathcal{R}}(k)} = 16 \epsilon_V(k_0) \Big( \frac{k}{k_0}\Big)^{4\epsilon_V-2\eta_V} \,\frac{\mathcal{T}_{T}(k)}{\mathcal{T}_{\mathcal{R}}(k)} \,. \label{TtoSrat}\ee Thus we see that for non-Bunch-Davis initial conditions the single-field slow roll consistency condition are modified to
\be r(k_0)= -8n_T(k_0)\,\Bigg[\frac{\mathcal{T}_{T}(k_0)}{\mathcal{T}_{\mathcal{R}}(k_0)} \Bigg] \label{consiscon}\ee and the standard consistency condition is \emph{not} fulfilled unless the transfer functions for curvature and tensor perturbations coincide at the pivot point $k_0$, which is obviously unlikely since the potentials for scalar and tensor perturbations are different and the pivot scale is arbitrary.

It remains to find $\mathcal{T}_\alpha(k)$.
The mode equation (\ref{modi}) with  $W (\eta)$ given by (\ref{Vpot}) can now be written as (we now drop the label $\alpha$ to avoid cluttering the notation)
\be  \Big[\frac{d^2}{d\eta^2}+k^2-\frac{\nu^2 -1/4}{\eta^2}\Big]S(k;\eta) = {V}(\eta) S(k;\eta) \,, \label{inteq} \ee and is  converted into an integral equation via the retarded Green's function $G_k(\eta,\eta')$ obeying
\be \label{green}
\left[\frac{d^2}{d\eta^2}+k^2-\frac{\nu^2-\frac14}{\eta^2}
 \right]G_k(\eta,\eta') = \delta(\eta-\eta') ~~;~~ G_k(\eta,\eta') =0~
\mathrm{for}~\eta'>\eta \,.
\ee
This Green's function  is given by
\be \label{Gret}
G_k(\eta,\eta') = i \left[g_\nu(k;\eta) \; g^*_\nu(k;\eta')-
g_\nu(k;\eta') \; g^*_\nu(k;\eta) \right] \Theta(\eta-\eta') \quad ,
\ee
where $ g_\nu(k;\eta) $ is given by eq.(\ref{BDmodes}).

We are interested in obtaining the power spectra for wavelengths of cosmological relevance today which crossed the Hubble radius during slow roll inflation. These modes were deep inside the Hubble radius during the fast roll stage and we take these to be described by the asymptotic behavior of Bunch-Davies modes $ \simeq e^{-ik\eta}/\sqrt{2k}$ for $-k\eta \gg 1$.

The solution of (\ref{inteq}) with boundary conditions corresponding to Bunch-Davies modes deep inside the horizon during the fast roll stage obeys the following Lippman-Schwinger integral equation familiar from scattering theory,
\be
S(k;\eta)= g_\nu(k;\eta) +
\int^{0}_{\eta_i} G_k(\eta,\eta') \; {V}(\eta') \;
S(k;\eta') \; d\eta' \;. \label{ecuaforS}
\ee

With the Green's function given by (\ref{Gret}) this solution can be written as
\be S(k;\eta) = A_k(\eta)g_\nu(k;\eta)+B_k(\eta)g^*_\nu(k;\eta) \,, \label{sola} \ee where
\bea  A_k (\eta)  & = & 1+  i \int^{\eta}_{\eta_i}
   {V}(\eta')\,g^*_\nu(k;\eta') \,S(k;\eta')  \, d\eta' \label{aofketa1} \\
 B_k(\eta)  & = & -  i \int^{\eta}_{\eta_i}
   {V}(\eta')\,g_\nu(k;\eta') \,S(k;\eta')  \, d\eta' \,.\label{bofketa1}\eea
 In ref.\cite{lellodS} it is shown that the $\eta$ dependent coefficients $A_k (\eta);B_k (\eta)$ obey
 \be \frac{d}{d\eta}\Bigg( |A_k(\eta)|^2-|B_k(\eta)|^2 \Bigg)=0 \label{const}\ee which by dint of the initial conditions at $\eta_i$ lead to the $\eta$-independent condition
 \be |A_k(\eta)|^2-|B_k(\eta)|^2 =1\,. \label{condiAB}\ee
 Since the potentials vanish for $\eta > \eta_{sr}$, the solution of the mode equations during the slow roll stage is given by
  \be S(k;\eta) = A_k g_\nu(k;\eta)+B_k g^*_\nu(k;\eta) ~~;~~\mathrm{for}~~\eta > \eta_{sr} \,, \label{solaSR} \ee namely of the form given by (\ref{Ssr}) with   the Bogoliubov coefficients   being the  solutions of the integral equations
\bea  A_k  & = & 1+  i \int^{\eta_{sr}}_{\eta_i}
   {V}(\eta')\,g^*_\nu(k;\eta') \,S(k;\eta')  \, d\eta' \label{aofketa} \\
 B_k  & = & -  i \int^{\eta_{sr}}_{\eta_i}
   {V}(\eta')\,g_\nu(k;\eta') \,S(k;\eta')  \, d\eta' \, \label{bofketa}\eea where the potentials for curvature and tensor perturbations are given by (\ref{curvapot},\ref{Vcurva}) and (\ref{tensorpot},\ref{vtensor}) respectively.
  The quantity $|B_k|^2$ has the interpretation of the number of Bunch-Davies particles created by the potential during the fast roll stage.

  Writing $S_{\alpha}$ as in eqn. (\ref{sola}) one obtains a coupled set of integral equations for the Bogoliubov coefficients and,  following ref.\cite{lellodS}, these can be written as a set of coupled differential equations which must be solved numerically in general. Ref.\cite{lellodS} provides an analysis of the behavior of the Bogoliubov coefficients in the long-wavelength limit, valid for modes that are superhorizon well before the onset of the slow roll stage. These modes remain outside the current Hubble radius and are of no observational significance today. Instead, we focus on modes that are deep within the Hubble radius during the fast roll stage and cross during the slow roll stage.

  The integral equations (\ref{aofketa},\ref{bofketa}) can be solved formally  as a Born series from the iterative solution of (\ref{ecuaforS}), namely
 \be S(k;\eta)= g_\nu(k;\eta) +
\int^{0}_{\eta_i} G_k(\eta,\eta') \; {V}(\eta') \;
g_\nu(k;\eta') \; d\eta' +\cdots \;, \label{Sborn}
\ee leading to the Born approximation for the Bogoliubov coefficients,
 \bea A_k & = &  1 + i \int^{\eta_{sr}}_{\eta_i}  {V}(\eta')\,|g_\nu(k;\eta')|^2 d\eta'+\cdots \label{Aborn}\\ B_k & = &  - i \int^{\eta_{sr}}_{\eta_i}  {V}(\eta')\,\big(g_\nu(k;\eta')\big)^2 d\eta'+\cdots \label{Bborn}\eea where we have used that $ {V}(\eta') =0 $ for $\eta > \eta_{sr}$.

   Progress can be made by recognizing that we are interested in wavevectors that have crossed the horizon during slow roll since those are of cosmological relevance today, therefore these wavevectors are deep within the Hubble radius during the fast roll stage $\eta_i \leq \eta \leq \eta_{sr}$. The mode functions $g_\nu(k;\eta) \propto 1/\sqrt{k}$ for wavevectors deep inside the horizon,  and from the expression for the potentials (\ref{curvapot},\ref{tensorpot}) and conformal time (\ref{etafx}) we recognize that ${V}_{\mathcal{R},T} \propto H^2_{sr}$ and $\eta \propto 1/H_{sr}$ therefore we expect that $A_k -1 ~;~B_k \propto H_{sr}/k$ suggesting that the lowest order Born approximation is reliable for wavectors that cross the horizon after $\eta_{sr}$, namely $k/H_{sr} > 1$ since we have normalized the scale factor so that $a(t_{sr})=1$. This expectation will be quantified and confirmed below. Furthermore to leading order in $\epsilon_V$ we will set $\nu = 3/2$ in the mode functions in (\ref{Aborn},\ref{Bborn}) with (again up to a phase)
   \be g_{3/2}(k;\eta)= -\frac{1}{\sqrt{2k}}\,e^{-ik\eta}\Big[1-\frac{i}{k\eta}\Big]\,. \label{g32}\ee

   To leading order in $\epsilon_V$ and in the Born approximation, namely linear order in the potentials ${V}$ we find
   \be \mathcal{T}_{\alpha}(k) = 1+ \frac{1}{k}\int_{\eta_i}^{\eta_{sr}} \,{V}_{\alpha}(\eta) \Bigg[ \frac{2\cos(2k\eta)}{k\eta}+\sin(2k\eta)\Big(1-\frac{1}{k^2\eta^2}\Big)\Bigg]\,d\eta \,.\label{Tis}\ee The potentials ${V}_\alpha$ have dimensions of $H^2_{sr}$ and $\eta$ has dimensions of $1/H_{sr}$ therefore it is convenient to define the dimensionless functions of the variable $x$ introduced in eqn. (\ref{xoft})
   \be \tilde{\eta}(x) \equiv H_{sr} \eta(x) \,,   \label{tileta}\ee where $\eta(x)$ is given by (\ref{etafx}) and
   \be \widetilde{{V}}_\alpha (x) = \frac{{V}_\alpha(\eta(x))}{H^2_{sr}}\,, \label{tipots}\ee along with the dimensionless ratio
   \be q = \frac{k}{H_{sr}}\,, \label{q} \ee in terms of which we find to leading order in the Born approximation
    \be \mathcal{T}_{\alpha}(k) = 1+  {D}_{\alpha}(q) \label{defD}\ee with

  \be  {D}_\alpha(q)=  \frac{1}{q}\,\Big(\frac{12}{\epsilon_V} \Big)^{1/6}\, \int_{x_i}^{x_{sr}} \,\frac{\widetilde{{V}}_{\alpha}(x)}{(1-x^6)^{1/3}} \Bigg[ \frac{2\cos(2q\tilde{\eta}(x))}{q\tilde{\eta}(x)}+\sin(2q\tilde{\eta}(x))\Big(1-\frac{1}{q^2\tilde{\eta}^2(x)}\Big)\Bigg]\,dx \,.\label{Dofq}\ee  The ratio $q$ has a simple interpretation: assuming that during slow roll the Hubble parameter does not vary appreciably, namely $H\simeq H_{sr}(1+\mathcal{O}(\epsilon_V))$ at least during the range of the slow roll regime when wavevectors of relevance today crossed the Hubble radius, a comoving wavevector $k$ corresponding to a physical scale that crosses the Hubble radius when the scale factor is $a_\star$ is given by $k = a_\star H_{sr}$ therefore $q=k/H_{sr} = a_\star$. Since we have normalized $a(t_{sr})=a_{sr}=1$ at the beginning of slow roll, values of $q >1$ correspond to physical wavelengths that cross the Hubble radius during the slow roll stage. If the slow roll stage of inflation lasts about 60 e-folds the wavelengths of relevance today crossed out of the Hubble radius during the first few e-folds after the beginning of slow roll and modes with $q = a_\star > 1$ are of cosmological relevance today.

\begin{figure}[h!]
\includegraphics[height=3.0in,width=3.2in,keepaspectratio=true]{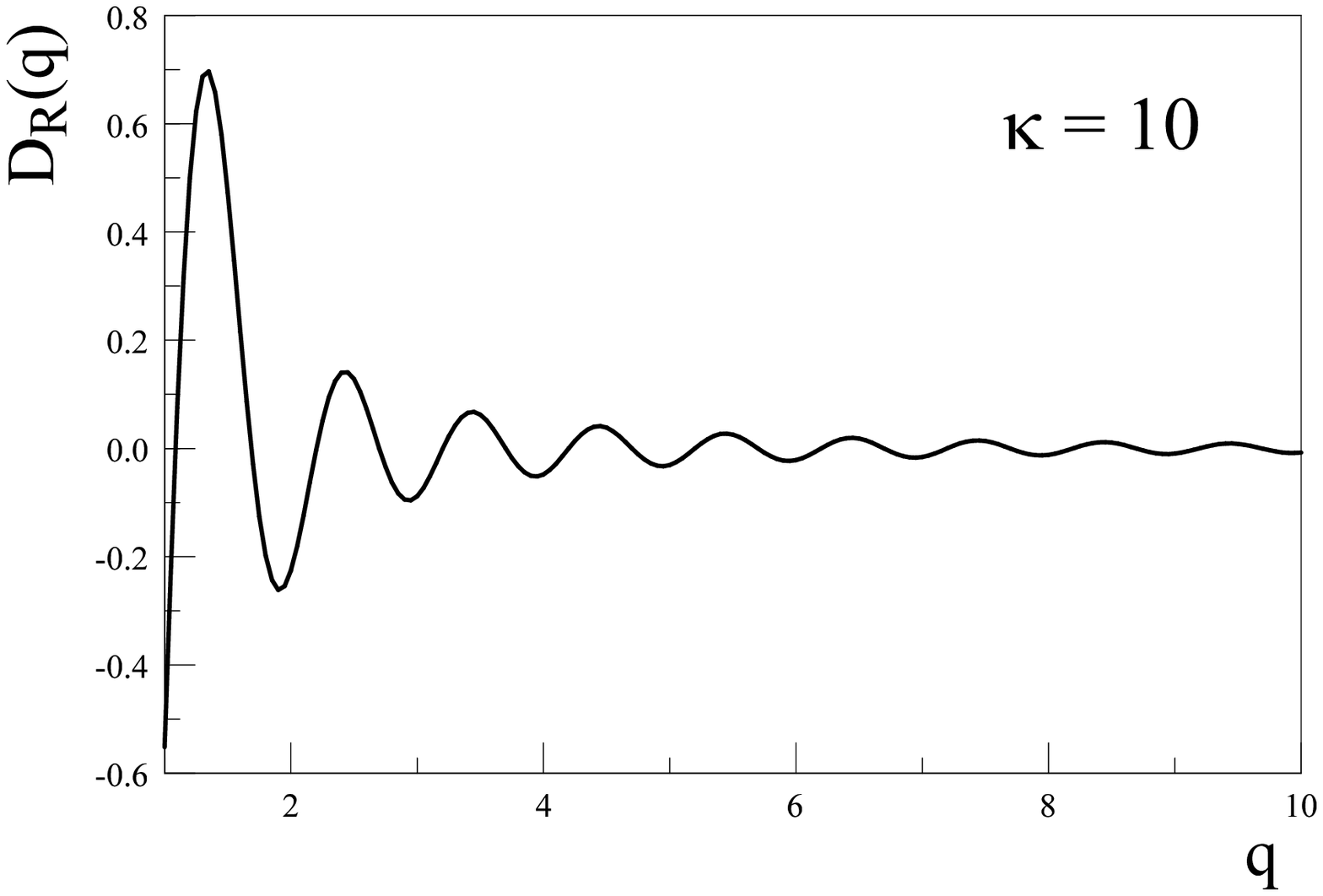}
\includegraphics[height=3.0in,width=3.2in,keepaspectratio=true]{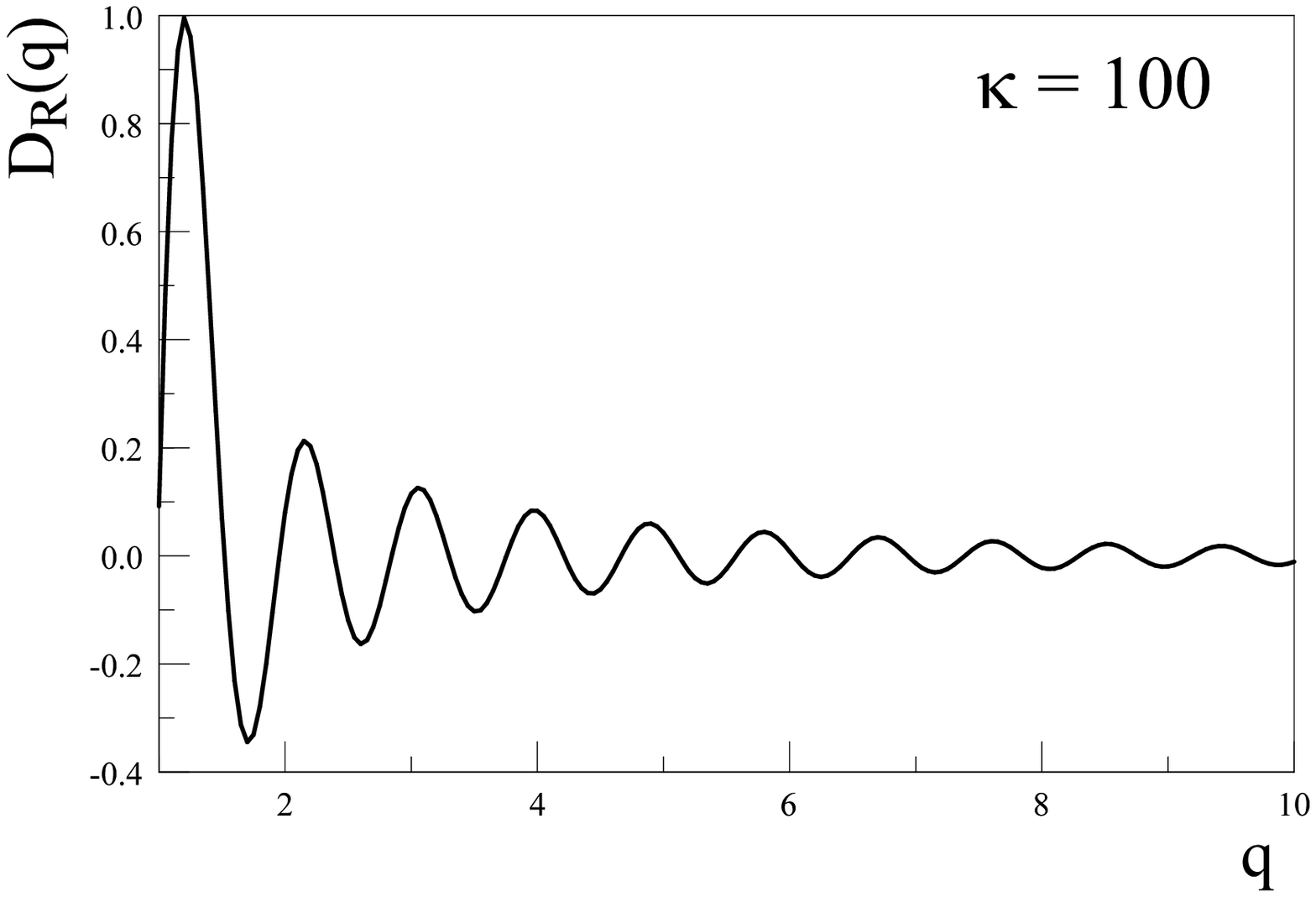}
\caption{$D_{R}(q)$   vs. $q=k/H_{sr}$ for $\kappa=10;100~;~\epsilon_V=0.008~;~\eta_V=-0.010$.  }
\label{fig:DR}
\end{figure}

\begin{figure}[h!]
\includegraphics[height=3.0in,width=3.2in,keepaspectratio=true]{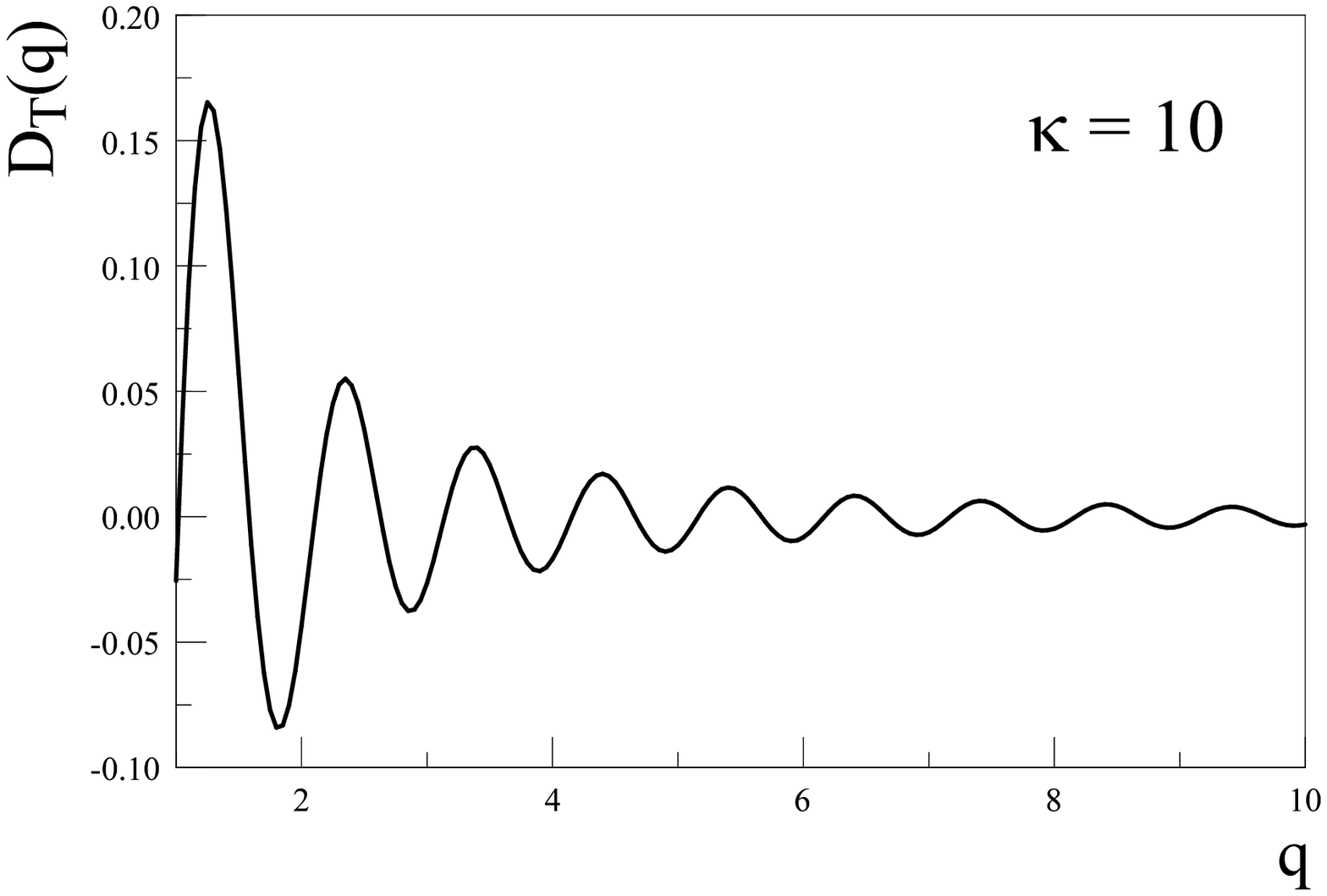}
\includegraphics[height=3.0in,width=3.2in,keepaspectratio=true]{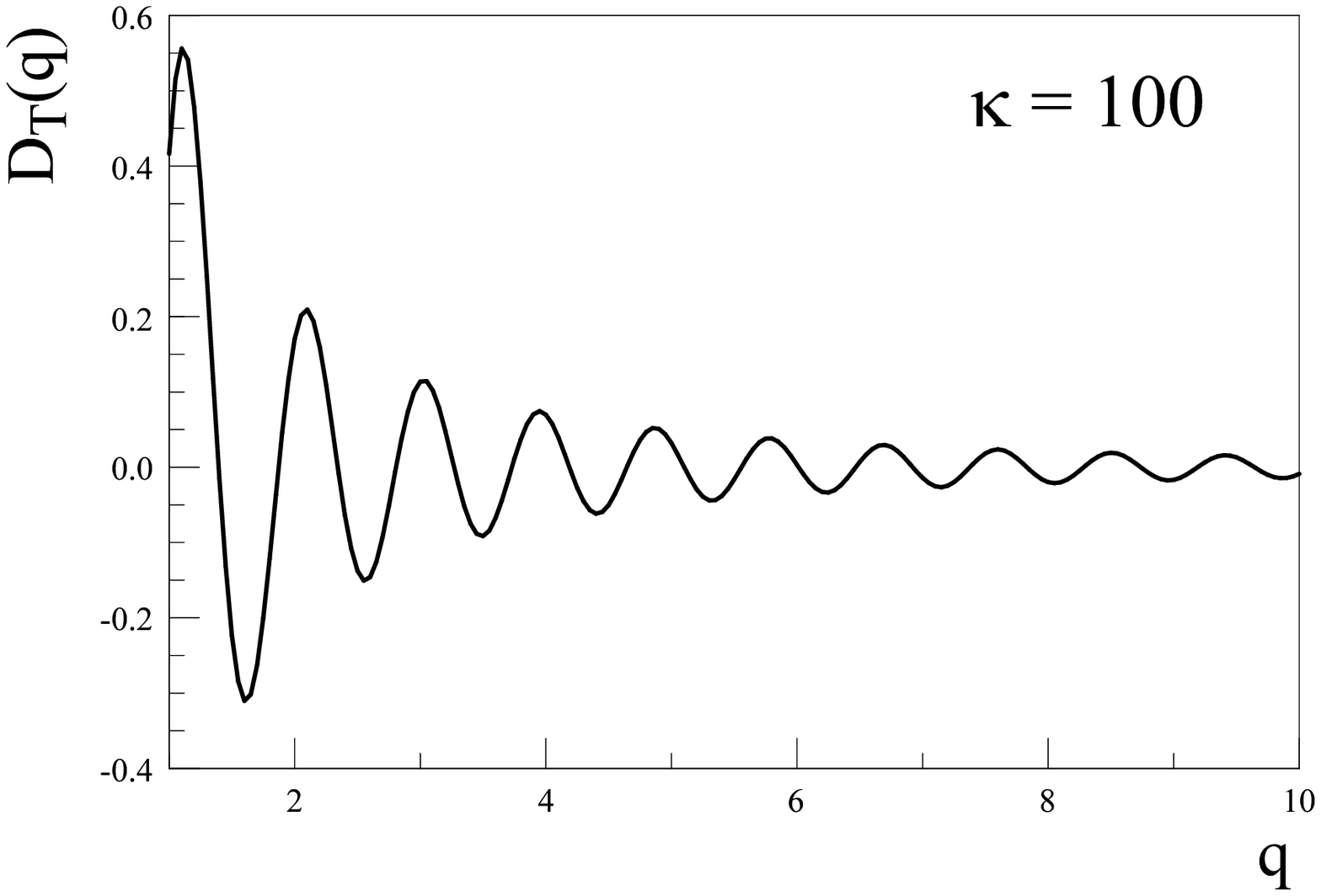}
\caption{  $D_T(q)$ vs. $q=k/H_{sr}$ for $\kappa=10;100~;~\epsilon_V=0.008~;~\eta_V=-0.010$.  }
\label{fig:DT}
\end{figure}

  Figs.(\ref{fig:DR},\ref{fig:DT}) show $D_R(q)$ and $D_T(q)$ for $\kappa =10,100$ for $ \epsilon_V=0.008, \eta_V=-0.01$. It  is clear that the Born approximation is reliable for $q>1$.

Although the Born approximation breaks down for $q<1$ these figures confirm  that the power spectra are suppressed at small $q$, as argued in ref.\cite{boyan3,hectordestri}, and feature oscillations of \emph{the same frequency } determined by the Hubble scale during slow roll inflation as revealed by the figures (\ref{fig:DR},\ref{fig:DT}). The observation of oscillations in the tensor to scalar ratio has been noted in other models, specifically in the double inflation model of ref.\cite{staro3} in which two distinct fields lead to separate periods of inflation leading to a mild oscillatory behavior in the period of transition.

Oscillatory behavior of the curvature and tensor power spectra and power suppression at small $k$  was also observed in ref.\cite{schwarz2} where a numerical integration of the mode equations from a kinetic dominated initial state was performed for the specific potential $\lambda \phi^4$. Although it is not straightforward to compare the scales, the discussion in this reference suggests that the oscillatory behavior is seen in modes that are very near horizon crossing and taper-off for larger values. This seems to be in agreement with our results that display oscillations for $q \simeq 1$ (namely $k \simeq H_{sr}$) but fall off as $\propto 1/q$ for $q >>1$.

 We emphasize that the results presented above depend solely on $\kappa, \epsilon_V;\eta_V$ but not on a specific realization of the inflationary potential, therefore are \emph{universal} in this sense.

\begin{figure}[h!]
\includegraphics[height=3.2in,width=3.2in,keepaspectratio=true]{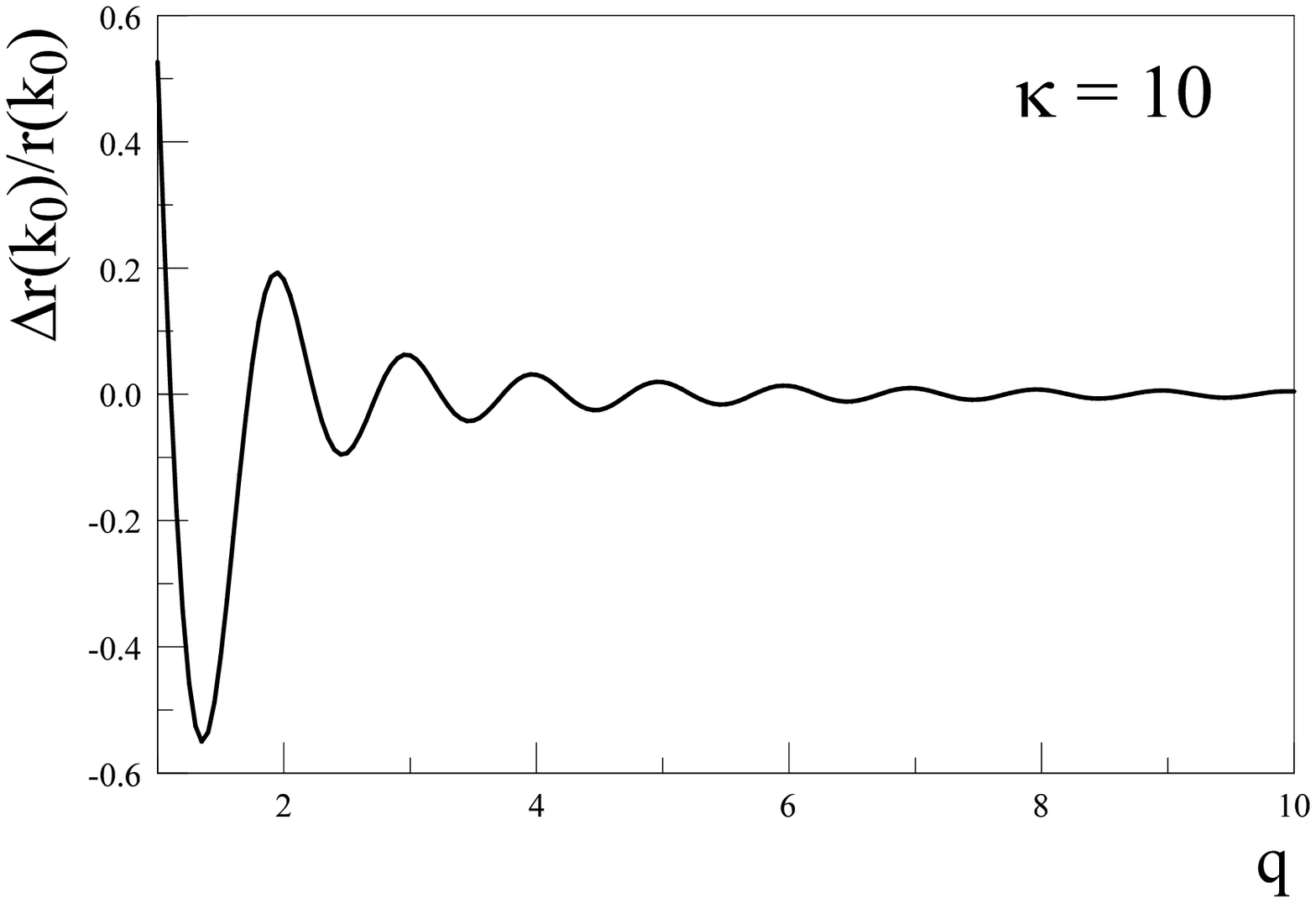}
\includegraphics[height=3.2in,width=3.2in,keepaspectratio=true]{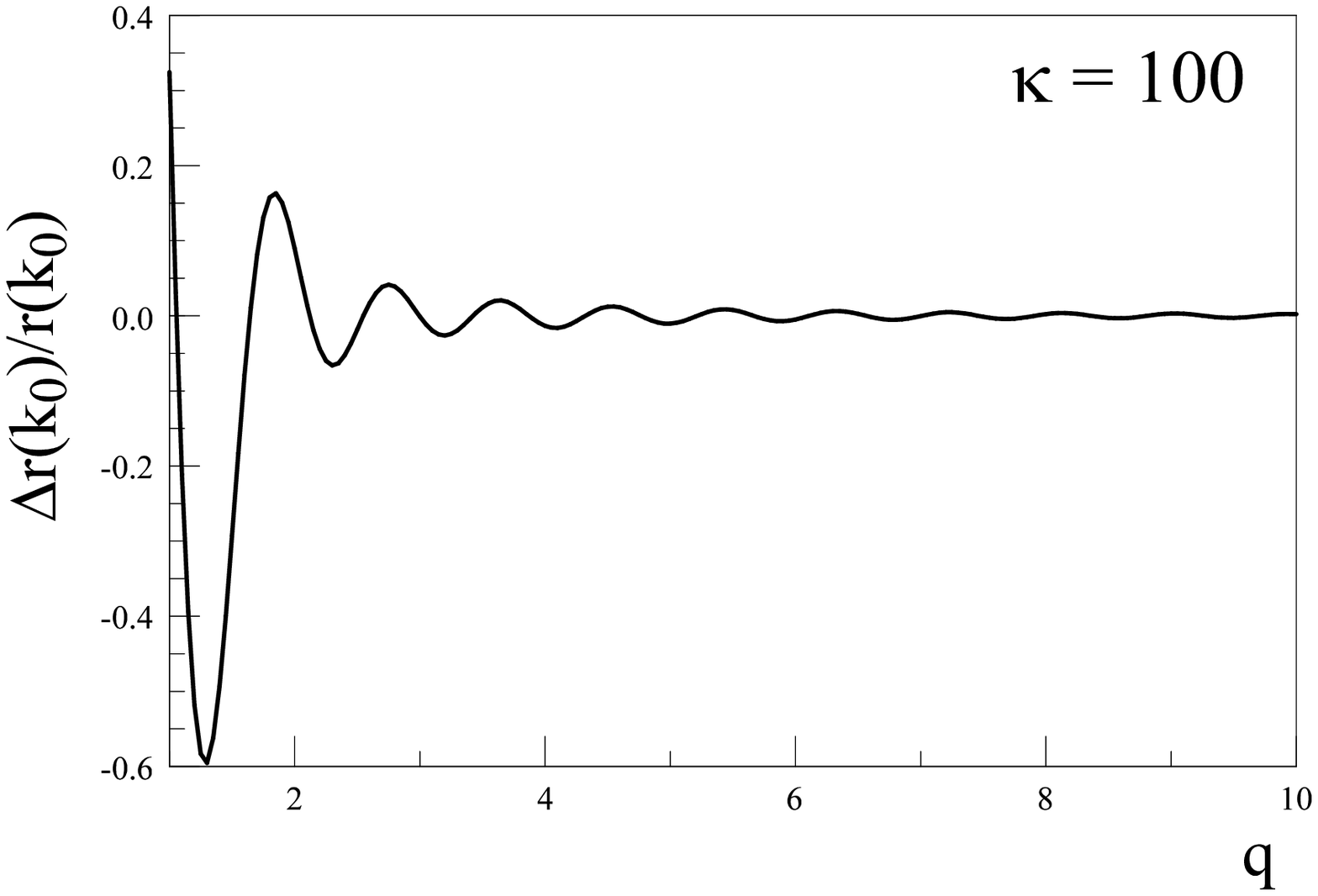}
\caption{$\Delta r(k_0)/r(k_0) = D_{T}(q)-D_R(q)$   vs. $q=k_0/H_{sr}$ for $\kappa=10;100~;~\epsilon_V=0.008~;~\eta_V=-0.010$.  }
\label{fig:difference}
\end{figure}

The relative change in the tensor to scalar ratio to leading order in the Born approximation is given by
\be \frac{\Delta r (k_0)}{r(k_0)} = D_T(q)-D_R(q) \label{deltar}\ee where $q = k_0/H_{sr}$ and $k_0$ is the pivot scale. This relative change  is displayed in fig.(\ref{fig:difference}) for $\kappa = 10;100$ for $\epsilon_V= 0.008;\eta_V=-0.010$.

The oscillatory features in the corrections to the curvature and tensor power spectra agree qualitatively with the results obtained in refs.\cite{boyan3,schwarz2} (which were obtained for different specific realizations of the inflationary potential). In fact, these oscillatory features are quite robust: the fast roll stage itself is insensitive to the potential, provided that the potential is very flat as is the case for slow roll inflation; it is only the merging with the slow roll stage that is sensitive to the potentials, but   only through the slow roll parameters $\epsilon_V,\eta_V$ to leading order in the slow roll expansion. Thus different potentials that lead to the same slow roll parameters would yield the same type of behavior the functions $D_{\alpha}$.

\vspace{3mm}

\textbf{Correlation with suppression of low multipoles:}

The modification on the initial conditions of the mode functions during slow roll  imprinted from the
pre-slow roll stage that lead to the corrections to the tensor to scalar ratio also affect the low multipoles in the CMB as previously discussed in refs.\cite{lindefast,contaldi,boyan3,hectordestri,nicholson,schwarz2}. In these references, specific inflationary potentials were studied whereas the analysis above, to leading order in the slow-roll parameters, is quite general and   depends solely on $\kappa, \epsilon_V,\eta_V$. This allows us to study the modifications on the low multipoles in a more general manner in order to establish a correlation between features in the tensor to scalar ratio and the suppression of the low multipoles, in particular the quadrupole.

In the analysis that follows we neglect the contributions to the $C_l's$ from the integrated Sachs-Wolfe effect (the  source of  secondary anisotropies that could  most likely   affect the large scale anomalies in the CMB\cite{francis,rassat}). This is supported by a recent analysis that suggests that a low quadrupole remains statistically significant and more anomalous even after subtraction of the (ISW) effect (see the discussion in ref.\cite{rassat}).

The modifications from the non-Bunch-Davies initial conditions upon the temperature power spectrum are encoded in the transfer function $\mathcal{T}_R(k)$ and, to leading order in the Born approximation, by the correction $D_R(q)$ which is given by (\ref{Dofq}) for $\alpha=R$ and the potential is given by (\ref{Vcurva}). This is depicted in fig.(\ref{fig:curvapotentials}).

In the region of the Sachs-Wolfe plateau for $l \lesssim 30$ the matter-radiation transfer function can be set to one and, neglecting the contribution from the (ISW) effect through dark energy, the
$C_l's$ are given by
\be C_l = \frac{4\pi}{9} \int_0^\infty \frac{dk}{k}\,\mathcal{P}_{\mathcal{R}}(k)\,j^2_l[k(\eta_0-\eta_{lss})]\, \label{Cl}\ee where the power spectrum $\mathcal{P}_{\mathcal{R}}(k)$ is given by (\ref{transfer}) and
\be \eta_0-\eta_{lss} = \frac{1}{a_0 H_0} \int^1_{1/(1+z_{lss})} ~\frac{dx}{\Big[\Omega_r+\Omega_m x+ \Omega_{\Lambda} x^4\Big]^{1/2}} = \frac{3.12}{a_0 H_0} \label{etadis}\ee where we have used $z_{lss}=1100$ and the latest parameters reported by the Planck collaboration\cite{planck}. To leading order in the Born approximation we find the relative  correction to the $C_l$ from the initial conditions to be
\be \frac{\Delta C_l}{C_l} = \frac{\int_0^{\infty} dk k^{n_s-2}\,D_{{R}}(k)\,j^2_l(3.12 k/a_0H_0)}{\int dk k^{n_s-2}\, j^2_l(3.12 k/a_0H_0)}\,. \label{delcl}\ee In particular, taking $n_s =1$ the corrections to the multipole $l$  are\footnote{This expression corrects an overall normalization in the second reference in\cite{boyan3}.}
\be \frac{\Delta C_l}{C_l} = 2l(l+1) \int_0^{\infty} \frac{dq}{q}  \,D_{{R}}(q)\,j^2_l\Big[\frac{3.12}{a_e} q\Big] \, \label{delc2}\ee where
\be a_e = \frac{a_0 H_0}{H_{sr}} \label{aexit}\ee is the value of the scale factor when the physical scale  corresponding to the Hubble radius today crossed the Hubble radius during slow roll inflation (with the normalization $a(t_{sr})=a_{sr}=1$ at the onset of slow roll).

The prominent oscillations in $D_{{R}}(q)$ do not yield oscillatory features in the ratio $\Delta C_2/C_2$ as a function of $a_e$. This result can be seen by combining (\ref{Dofq}) with (\ref{delc2}), which leads to
\be \frac{\Delta C_2}{C_2} = \Big(\frac{12}{\epsilon_V} \Big)^{1/6}\, \int_{x_i}^{x_{sr}} \,\frac{\widetilde{{V}}_{R}(x)}{(1-x^6)^{1/3}} \Psi(\tilde{\eta}(x))\,dx\label{psifun} \ee
where the function $\Psi(\tilde{\eta}(x))$ has been studied in    the second reference in\cite{boyan3} (see appendix of this reference), this function is non-oscillatory and  positive for $\tilde{\eta} < 0$. Therefore for ${V}_R < 0$ it follows that $\Delta C_2$ is non-oscillatory and negative as a function of $a_e$.

\begin{figure}[h!]
\begin{center}
\includegraphics[height=2.5in,width=3.0in,keepaspectratio=true]{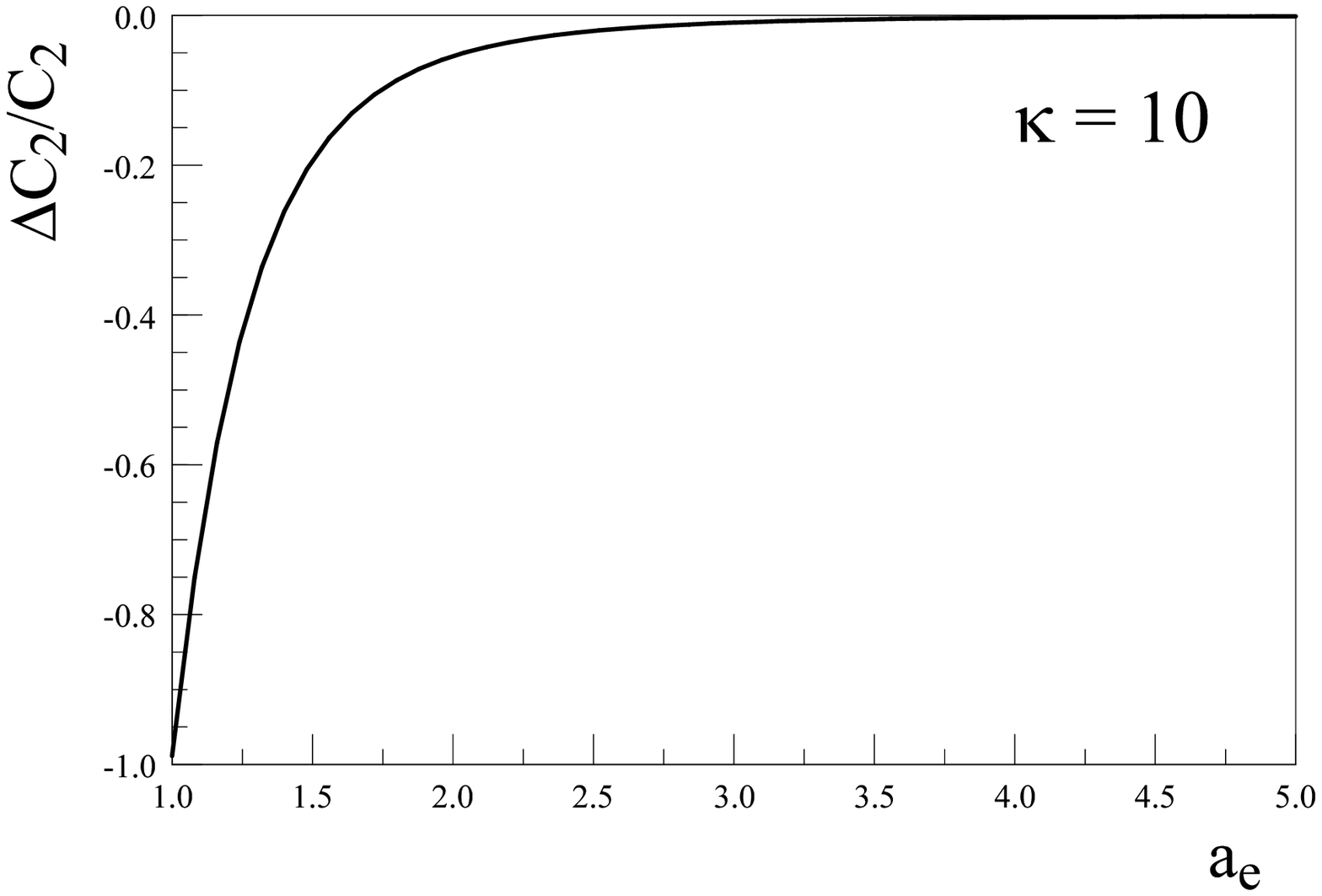}
\includegraphics[height=2.5in,width=3.0in,keepaspectratio=true]{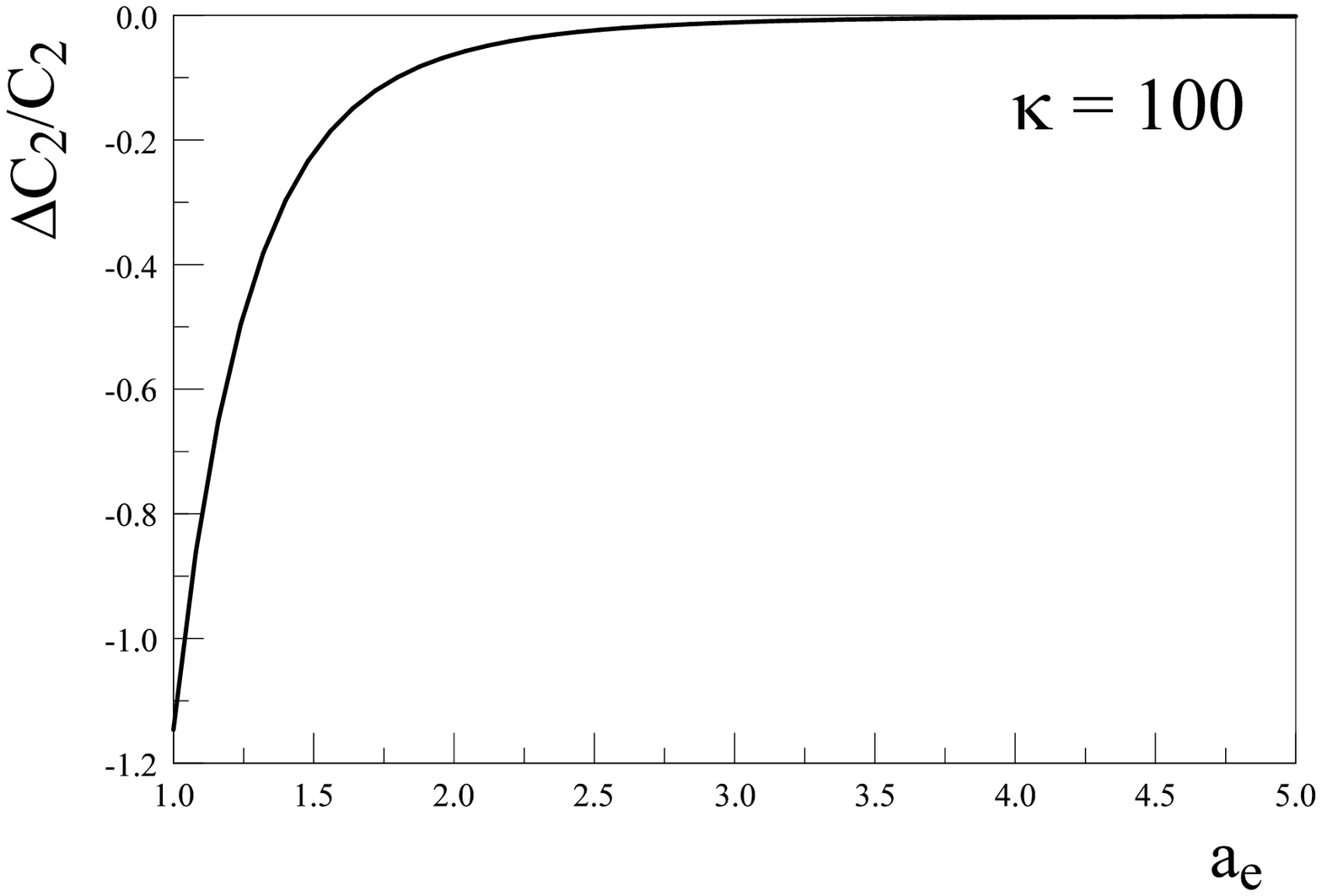}
\caption{$\Delta C_2/C_2$   vs. $a_e$  for $\kappa=10;100~;~\epsilon_V=0.008~;~\eta_V=-0.010$.  }
\label{fig:del2}
\end{center}
\end{figure}

The relative change in the quadrupole and octupole are shown in fig. (\ref{fig:del2}) for $\kappa=10;100$ and $\epsilon_V=0.008;\eta_V=-0.010$.

 These figures reveal that a $5-10\%$ suppression of the quadrupole, as reported by the Planck collaboration\cite{planck2} can be accounted for if $a_e \simeq 2-3$.   These  results for $\Delta C_2/C_2$ are qualitatively similar to those found in ref.\cite{boyan3} for a specific potential and different values of $\epsilon_V,\eta_V$ (as well as the argument of the Bessel function) and different initial conditions.\footnote{There is also a normalization error in the second reference in\cite{boyan3}.}

%  We have also studied $\Delta C_l/C_l$ for $3 \leq l \leq 30$ in the Sachs-Wolfe plateau and found consistently that the higher multipoles are substantially suppressed. The typical changes are $\Delta C_l/C_l \ll \epsilon_V$ for $l> 3$ and, therefore, unobservable and indistinguishable from higher order corrections in $\epsilon_V$.

The function $j^2_2\big[\frac{3.12}z\big]/z$ features a  sharp peak at $z\simeq 1$ of width $\Delta z \simeq 1$, therefore the largest contribution to the integrand in (\ref{delc2}) for the quadrupole ($l=2$) arises from the region in $q$ centered at $q \simeq a_e$ of width $\Delta q \simeq a_e$. From fig. (\ref{fig:del2}) we see that for $2 \leq a_e \lesssim 4$ there is a suppression in the quadrupole in the range $0.05 \leq \Delta C_2/C_2 \lesssim 0.1-0.15$ which is approximately the  suppression reported by the  Planck collaboration\cite{planck2}. Translating this range to fig. (\ref{fig:difference}), we see that if the pivot scale $k_0$ is such that $ 2\lesssim q=k_0/H_{sr} \lesssim 6-7$ then the tensor to scalar ratio should display oscillations with a period $\simeq H_{sr}$ as a function of the pivot scale.

Therefore, if the total number of inflationary e-folds is   about the minimum for the scale corresponding to the Hubble radius today to have crossed the Hubble radius near the beginning of slow roll inflation, then the fast-roll stage would lead to a suppression of the quadrupole consistent with observations \emph{and} oscillations in the tensor to scalar ratio. These \emph{could} be observable if the wavelength corresponding to the pivot scale crosses the Hubble radius during slow roll just a few e-folds after the beginning of slow roll.

 Additionally, it has been pointed out in \cite{contaldi2} that relieving the tension between Planck and BICEP by the invocation of a running spectral index is statistically less preferential than a mechanism which would lead to a large scale power suppression. While ref.\cite{contaldi2} finds that a power suppression of $\simeq 35\%$ to be the best fit to Planck+BICEP data and we find a $\simeq 10\%$ suppression due to the fast roll stage, we emphasize that our analysis was only a leading order correction and further numerical effort would be needed to fully ascertain the effect of large scale power suppression due to a fast roll scenario.

\begin{figure}[h!]
\begin{center}
\includegraphics[height=2.5in,width=3.0in,keepaspectratio=true]{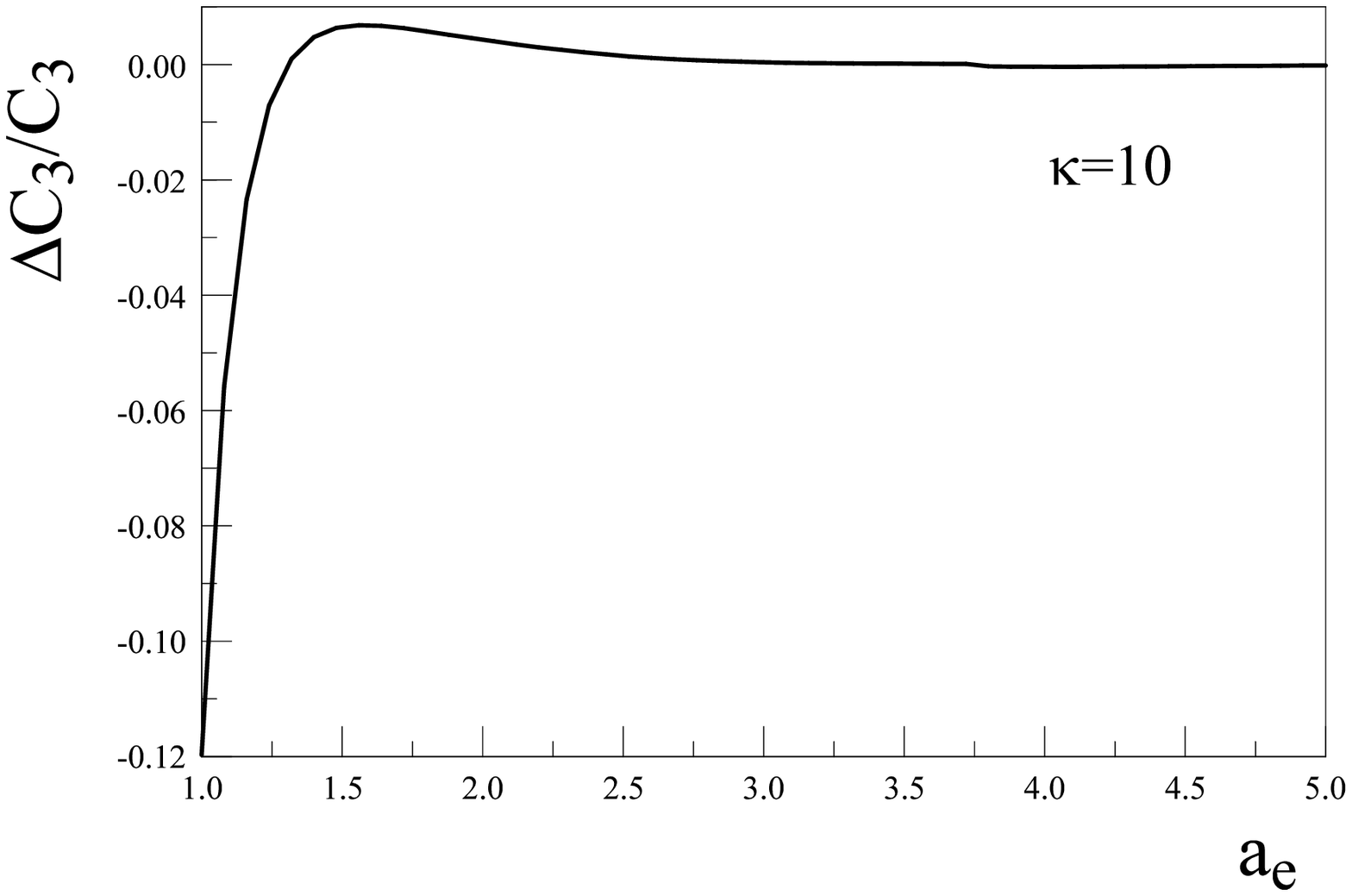}
\includegraphics[height=2.5in,width=3.0in,keepaspectratio=true]{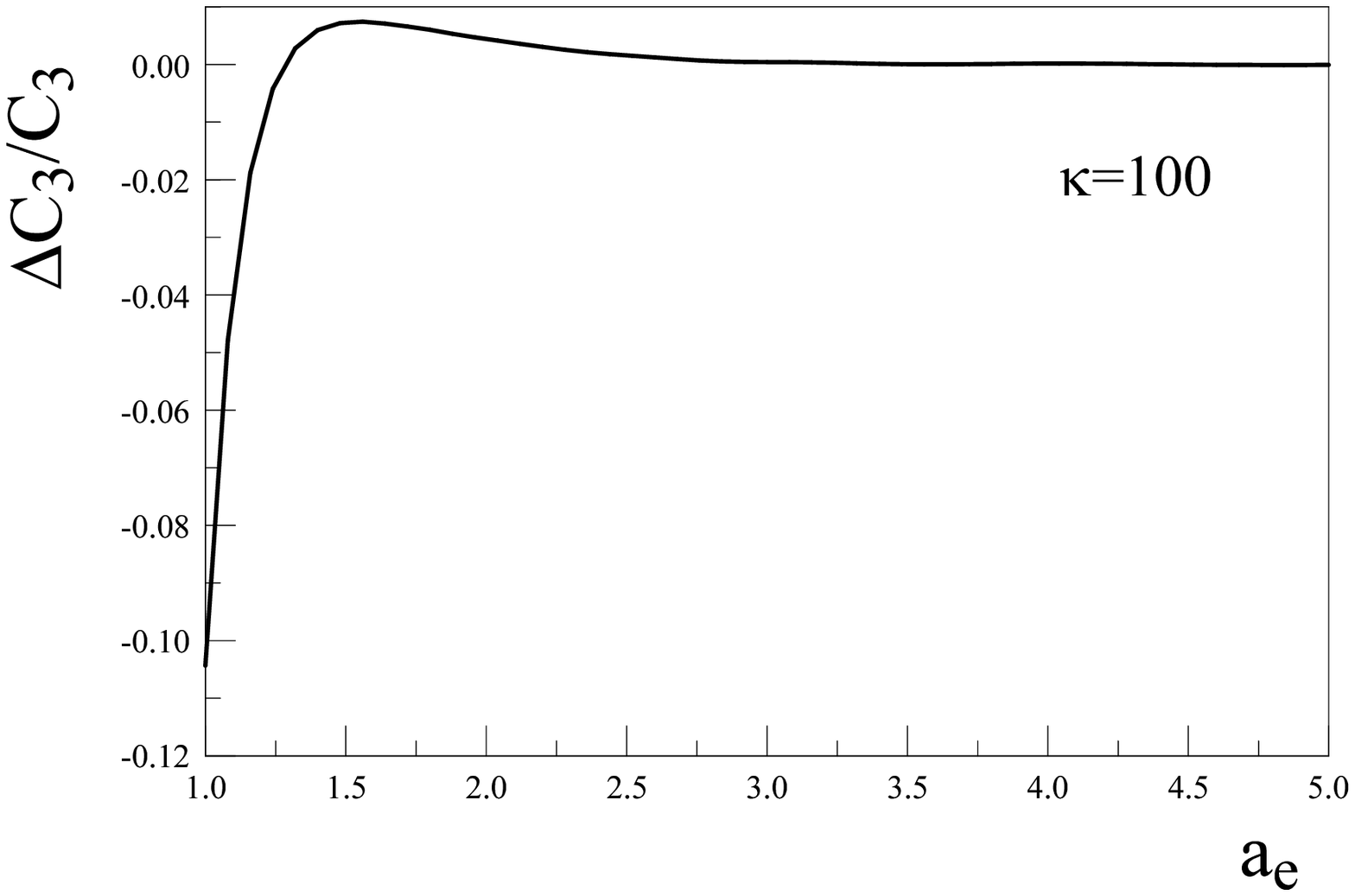}
\caption{  $\Delta C_3/C_3$ vs. $a_e$  for $\kappa=10;100~;~\epsilon_V=0.008~;~\eta_V=-0.010$.  }
\label{fig:del3}
\end{center}
\end{figure}

We have also studied $\Delta C_l/C_l$ for $3 \leq l \leq 30$ in the Sachs-Wolfe plateau and found consistently that the higher multipoles are not substantially suppressed with respect to $\Delta C_2/C_2$.
Fig. (\ref{fig:del3}) displays the corrections to the octupole, which are consistently at least an order of magnitude smaller than $\Delta C_2/C_2$ in the whole range ($a_e >1$), and displays a slight enhancement at $a_e \simeq 2$, a region where the quadrupole shows a suppression of $5-10\%$ which is consistent with Planck results\cite{planck2}; however, the amplitude of such enhancement is $\simeq \mathcal{O}(\epsilon_V)$. For $l\geq 3$ the typical changes $\Delta C_l/C_l \ll \epsilon_V$   and, therefore, unobservable and indistinguishable from higher order corrections in $\epsilon_V$.

\begin{figure}[h!]
\begin{center}
\includegraphics[height=2.5in,width=3.0in,keepaspectratio=true]{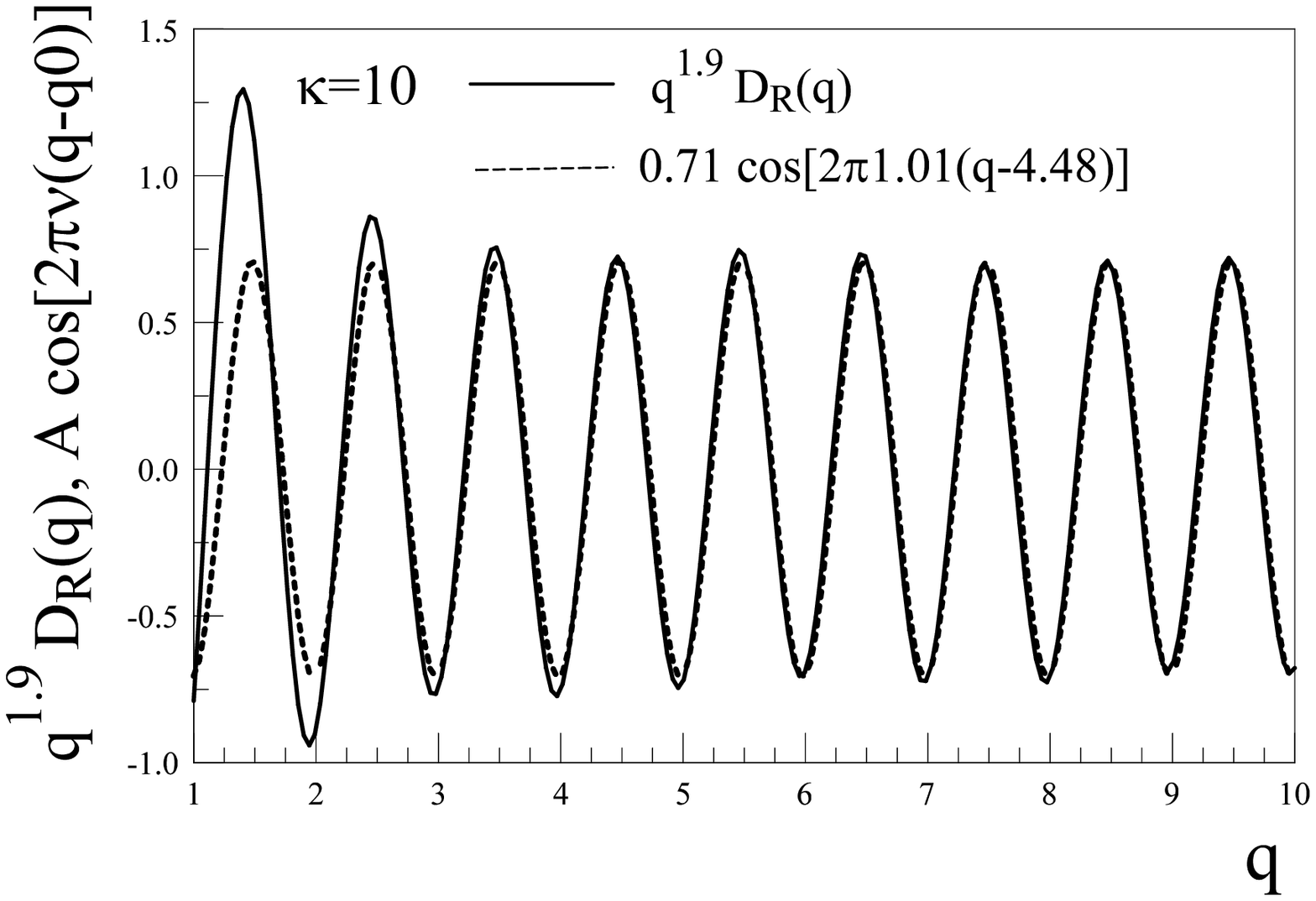}
\includegraphics[height=2.5in,width=3.0in,keepaspectratio=true]{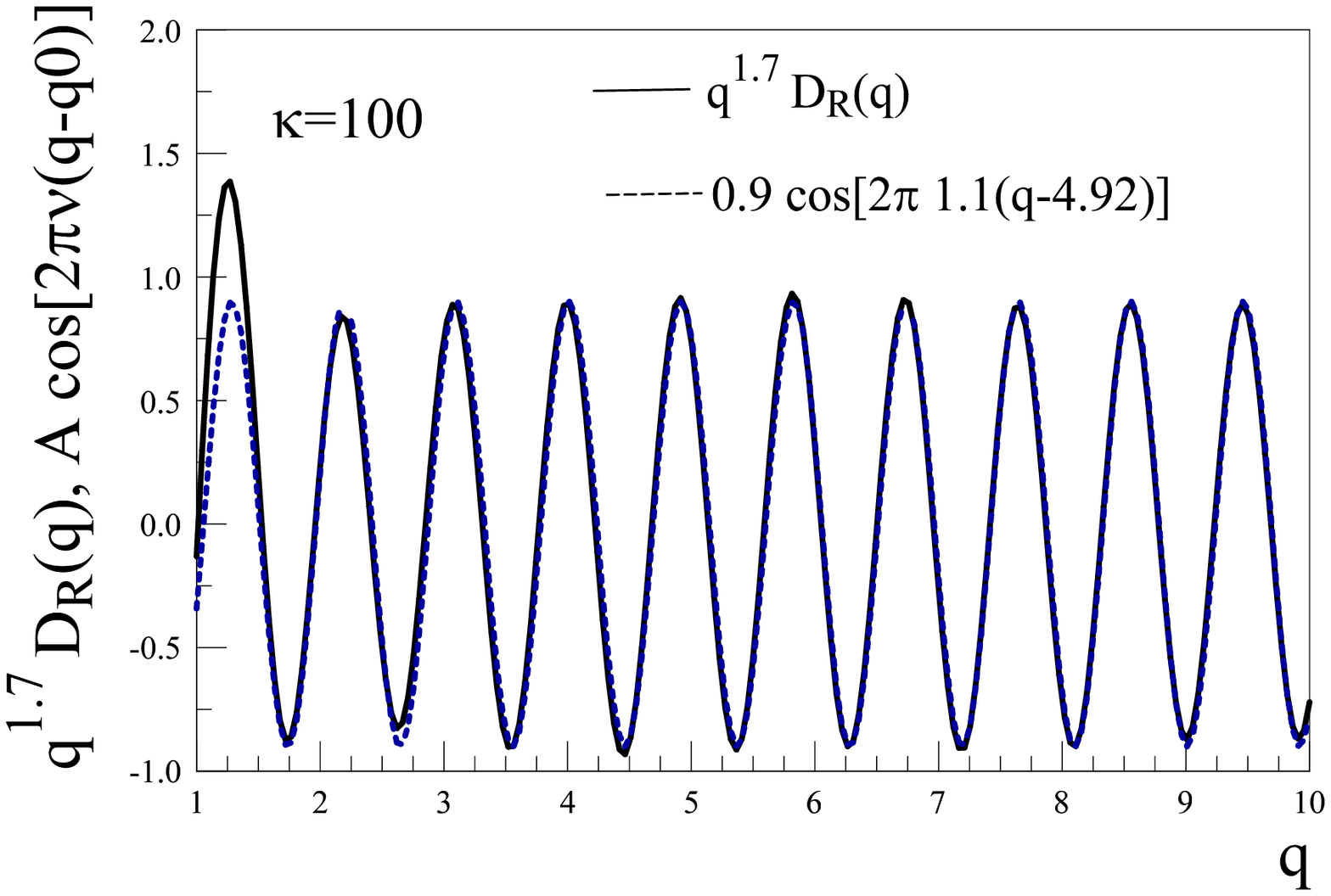}
\includegraphics[height=2.5in,width=3.0in,keepaspectratio=true]{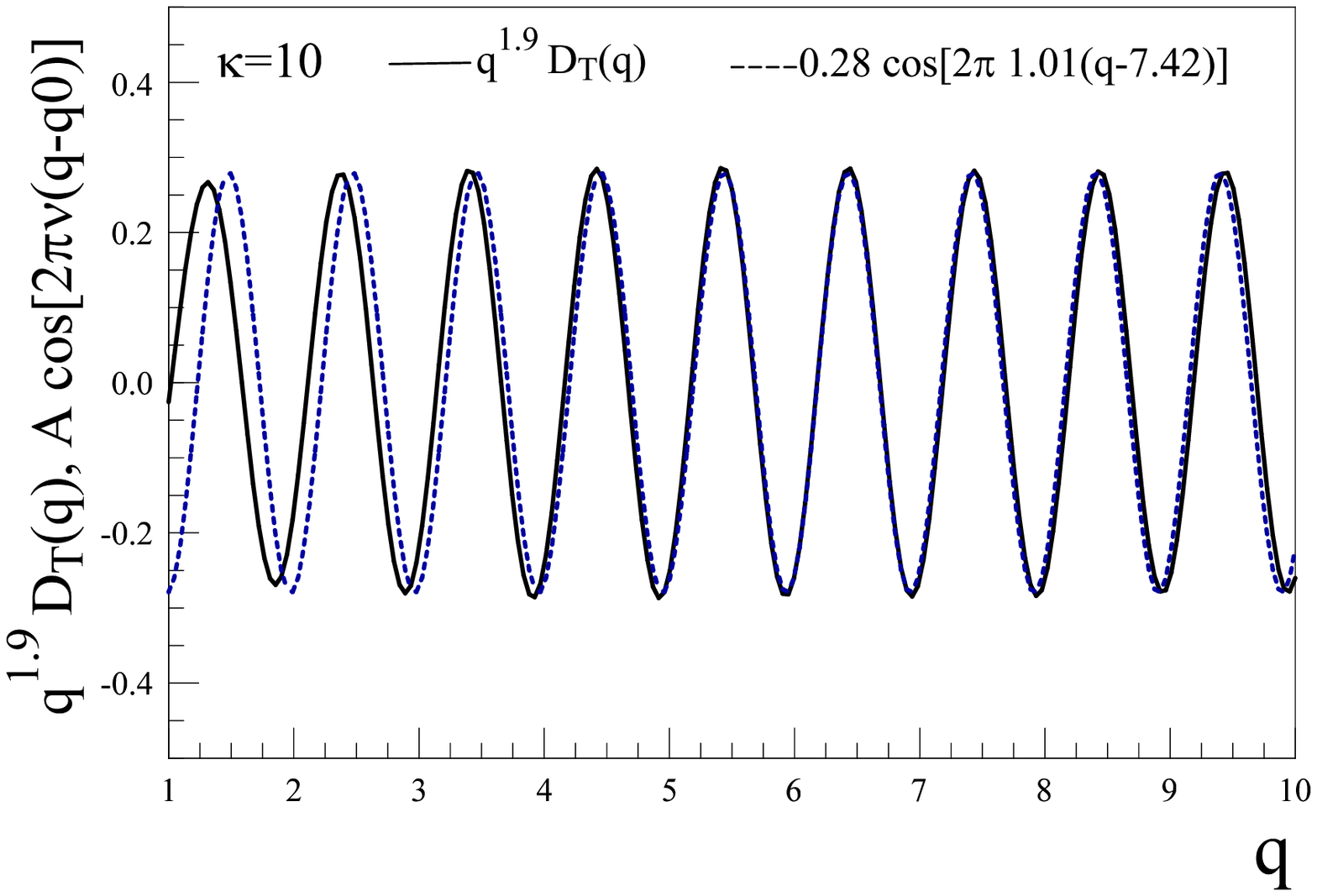}
\includegraphics[height=2.5in,width=3.0in,keepaspectratio=true]{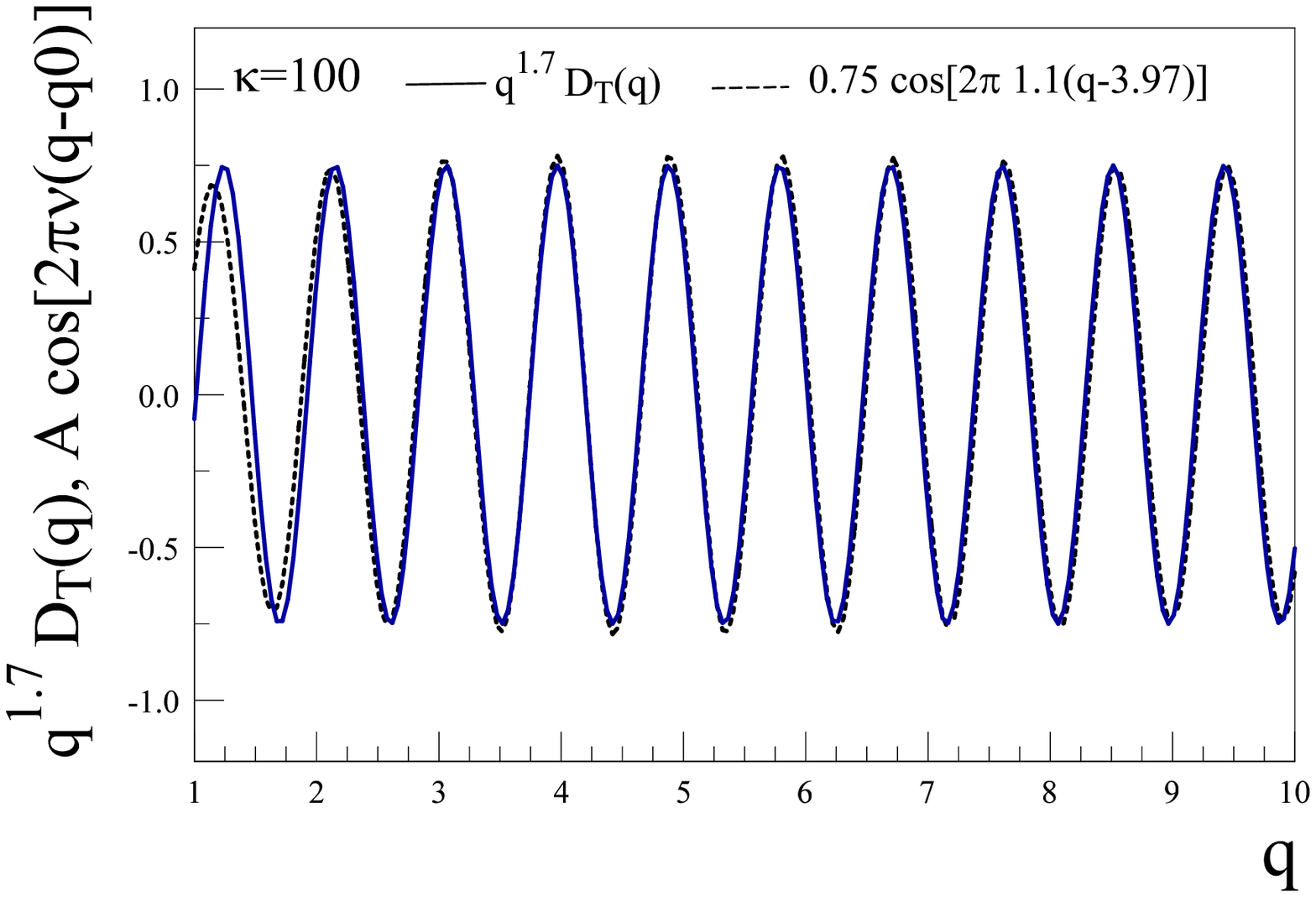}
\caption{Fits to $D_{R,T}(q)$ within the range $1\leq q \leq 10$  for $\kappa=10;100~;~\epsilon_V=0.008~;~\eta_V=-0.010$.  .}
\label{fig:fits}
\end{center}
\end{figure}

Recently a method to search for oscillatory features in power spectra was introduced\cite{meersper} that bears the promise of extracting important signatures from (CMB) data that could help to discern the effects of non-Bunch Davies initial conditions.   In order to search for features in (CMB) data it would be helpful to find a particular simple  form for the corrections that could be useful for analysis advocated in ref.\cite{meersper}. We find that while there is no simple fit to the damped oscillatory form of the corrections generally valid for all values of $q$, there are simple fits that are valid in a wide range of momenta of observational relevance. In particular,  within the wide interval $1.5 \lesssim q \lesssim 10-15$, the following form is a very accurate fit both for curvature and tensor perturbations
\be D_{\alpha}(q) =\frac{ A_\alpha(\kappa)}{q^{p(\kappa)}}\,\cos\big[2\pi\,\omega(\kappa) \,q+\varphi(\kappa) \big] \label{fit2} \ee where the amplitude, power, frequency and phase-shift are slowly varying functions of $\kappa$ within the wide range $3\lesssim \kappa \lesssim 100$. Remarkably we find that the power and the frequency are \emph{the same} for both types of perturbations, the power diminishes with $\kappa$ within the range $1.5\lesssim p(\kappa)\lesssim 2$ for $3\lesssim \kappa  \lesssim 100$ whereas $1 \lesssim \omega \lesssim 1.1$ within this range. These fits are shown in fig.(\ref{fig:fits}) for both curvature and tensor perturbations. Whereas both the power $p$ and frequency $\omega$ are the \emph{same} (at least within the accuracy of the numerical fit) for both curvature and tensor perturbations, they differ both in amplitude and phase-shifts.

This analysis allows us to provide a compact form for the curvature and tensor power spectra that includes the modifications from the pre-slow roll stage and is valid within a wide range of observationally relevant momenta for the low-l region of the (CMB):

\be  \mathcal{P}_R(k)    =    \mathcal{A}_R(k_0)\,\Big(\frac{k}{k_0} \Big)^{n_s-1}\Bigg[1+ A_R(\kappa) \Big( \frac{H_{sr}}{k}\Big)^{p(\kappa)}\, \cos\big[2\pi\,\omega(\kappa) \,\frac{k}{H_{sr}} +\varphi_R(\kappa) \big] \Bigg] \label{PRcorr} \ee
\be \mathcal{P}_T(k)    =    \mathcal{A}_T(k_0)\,\Big(\frac{k}{k_0} \Big)^{n_T}\Bigg[1+ A_T(\kappa) \Big( \frac{H_{sr}}{k}\Big)^{p(\kappa)}\, \cos\big[2\pi\,\omega(\kappa) \,\frac{k}{H_{sr}} +\varphi_T(\kappa) \big] \Bigg]\,. \label{PTcorr} \ee

Clearly there are further corrections to the Bunch-Davies part of the power spectra which are higher order in slow roll parameters $\epsilon_V,\eta_V$, however, these are non-oscillatory and cannot mask the oscillatory contributions in the brackets of these expressions.

\section{Summary, conclusions and further questions}
Motivated by the most recent results from the PLANCK collaboration\cite{planck,planck2,planck3} reporting statistically significant anomalies at large scales, in this article we study the corrections to the curvature and tensor power spectra from non-Bunch-Davies initial conditions resulting from a fast-roll stage prior to slow roll in single field inflation with canonical kinetic terms. We consider initial conditions in which the kinetic energy of the inflaton is larger than the potential energy, parametrized by the ratio
\be \frac{\dot{\Phi}^2_i}{2 V_{sr}} = \kappa \gg 1 \ee where $V_{sr}$ is the (fairly flat) inflaton potential consistent with slow roll. For a wide range of initial conditions with $  \kappa \lesssim 100$ the pre-slow roll, kinetic dominated stage lasts for about $2-3 $ e-folds with the inflationary stage beginning promptly within $\simeq 1$ e-folds, merging smoothly with the slow roll phase.

We have developed a program that yields the solution for the dynamics of the inflaton that interpolates between the fast and slow roll stages consistently in an expansion in $\epsilon_V,\eta_V$ which is
\emph{independent of the inflationary potential}. This approach relies on a separation of time scales and is valid for general inflationary potentials that are monotonic and can be described by a derivative expansion characterized by the slow roll parameters.

The fast roll stage modifies the potentials that enter in the equations of motion for the mode functions of curvature and tensor perturbations resulting in   non-Bunch-Davies initial conditions for the mode functions \emph{during slow roll}. The power spectra for curvature and tensor perturbations ($\alpha = R, T$) are modified to
\be \mathcal{P}_{\alpha}(k)=P^{BD}_{\alpha}(k)\mathcal{T}_\alpha(k) \ee where $\mathcal{T}_\alpha(k)$ are transfer functions determined by the non-Bunch-Davies initial conditions. These corrections entail a modification of the ``consistency conditions'' for the tensor to scalar ratio for single field inflation (with canonical kinetic term) to
\be r(k_0)= -8n_T(k_0)\,\big[ {\mathcal{T}_{T}(k_0)}/{\mathcal{T}_{\mathcal{R}}(k_0)} \big] \ee where $k_0$ is the ``pivot'' scale.

We obtain explicit expressions for the $\mathcal{T}_\alpha(k)$ in a Born approximation which is valid for all modes of observational relevance today; i.e. those that had crossed the Hubble radius during inflation within $1-2$ e-folds from the beginning of the slow roll stage. The modification of the power spectrum for curvature perturbations yields a suppression of the (CMB) quadrupole  consistent with the results from Planck\cite{planck2} if the modes corresponding to the Hubble radius today crossed the Hubble radius within a few ($\simeq 2-3$) e-folds from the beginning of slow roll,  suggesting that  a kinetic dominated pre-slow roll stage is a possible explanation of  the quadrupole suppression if the number of e-folds during slow roll inflation is the minimal required to solve the horizon problem $\simeq 60-62$. As discussed in \cite{contaldi2}, a large scale power suppression is a mechanism which could serve to relieve the tension between the Planck and BICEP experiments which would imply that a fast roll stage \emph{could} potentially serve as a mechanism to explain the seemingly conflicted results.

The suppression of the quadrupole is correlated with oscillatory features in the tensor to scalar ratio which \emph{could} be observable, again if the modes corresponding to the observed pivot scale crossed the Hubble radius a few e-folds   from the beginning of slow roll inflation.

A numerical fit to the power spectra valid for these wavevectors yields
\be  \mathcal{P}_R(k)    =    \mathcal{A}^{BD}_R(k_0)\,\Big(\frac{k}{k_0} \Big)^{n_s-1}\Bigg[1+ A_R(\kappa) \Big( \frac{H_{sr}}{k}\Big)^{p(\kappa)}\, \cos\big[2\pi\,\omega(\kappa) \,\frac{k}{H_{sr}} +\varphi_R(\kappa) \big] \Bigg]   \ee
\be \mathcal{P}_T(k)    =    \mathcal{A}^{BD}_T(k_0)\,\Big(\frac{k}{k_0} \Big)^{n_T}\Bigg[1+ A_T(\kappa) \Big( \frac{H_{sr}}{k}\Big)^{p(\kappa)}\, \cos\big[2\pi\,\omega(\kappa) \,\frac{k}{H_{sr}} +\varphi_T(\kappa) \big] \Bigg]   \ee remarkably with the same power $p(\kappa)$ and frequency $\omega(\kappa)$ for \emph{both} tensor and curvature perturbations, with $H_{sr}$ the Hubble scale during inflation and
\be 1.5 \lesssim p(\kappa) \lesssim 2 ~;~ \omega(\kappa) \simeq 1 \ee  and
\be 0.7  \lesssim A_R(\kappa) \lesssim 0.9 ~~;~~ 0.3 \lesssim A_T(\kappa) \lesssim 0.8   \ee
for the range $3\lesssim \kappa \lesssim 100$.

Perhaps these oscillatory features in the power spectra \emph{may be} extracted from (CMB) data with the implementation of the techniques advocated recently in ref.\cite{meersper}. The Bunch-Davis contribution to the power spectra will feature higher order corrections in $\epsilon_V,\eta_V$ which \emph{may be comparable in magnitude} to the corrections brought about by the fast roll phase, however, the distinguishing oscillatory features in the power spectra above cannot be confused with the non-oscillatory higher order slow roll corrections.

The results discussed above were obtained within the regime of validity of the Born approximation and to leading order in an expansion in $\epsilon_V,\eta_V$,  therefore there remains the question of possible corrections beyond this approximation, for which a more definitive answer would imply either extending the calculation to  higher orders in the Born series and the $\epsilon_V,\eta_V$ expansion of section (\ref{sec:match}) or a full numerical solution of the mode equations and the Friedmann equation. Both approaches imply a substantial and intensive numerical effort, an  endeavor that would be justified if the analysis of the (CMB) data yields hints of oscillatory behavior in broad agreement with the scales and general features  of the results of the leading order approximation described by the power spectra above. With the recent detection of primordial B-waves in ref. \cite{bicep2}, the proposal of scanning across pivot scales in an effort to observe the aforementioned oscillations in the tensor to scalar ratio is a potentially realistic future goal and, if such oscillations are detected, direct access to pre-inflationary information may be within the realm of plausibility.

\acknowledgments  L. L. and D.B.  acknowledge partial support from NSF-PHY-1202227. The authors thank P. D. Meerburg for an insightful conversation.

\end{document}